\documentclass[aps,prl,reprint,floatfix,amssymb,amsmath,preprintnumbers]{revtex4-2}
\usepackage{graphicx}
\usepackage{color}
\usepackage{boxhandler}
\usepackage{dcolumn}
\usepackage{ulem}
\usepackage{subfiles}

\usepackage[colorlinks=true,allcolors={apsBlue}]{hyperref}
\definecolor{apsBlue}{RGB}{50, 50, 150}

\newcommand{\ket}[1]{\ensuremath{| #1 \rangle}}

\begin{document}

\title{One second interrogation time in a 200 round-trip waveguide atom interferometer}
\preprint{LA-UR-22-20692}

\author{Hyosub Kim,$^{1,\ast,\dagger}$ Katarzyna Krzyzanowska,$^{1}$ K. C. Henderson,$^{1}$ C. Ryu,$^{1}$ Eddy Timmermans,$^{2}$ \& Malcolm Boshier$^{1,\ddag}$}
\affiliation{
 \mbox{$^{1}$MPA-Q, Los Alamos National Laboratory, Los Alamos, NM 87545,}\\
 \mbox{$^{2}$XCP-5, Los Alamos National Laboratory, Los Alamos, NM 87545}\\
}

\date{\today}

\begin{abstract}{
We report a multiple-loop guided atom interferometer in which the atoms make 200 small-amplitude round-trips, instead of one large single orbit. The approach is enabled by using ultracold $^{39}$K gas in 1D waveguide potential formed by a loosely focusing 1064~nm dipole trap and a magnetic Feshbach resonance that can tune the $s$-wave scattering length across zero to significantly reduce the atom loss from cold collisions. The momentum transfer is imparted by a Bragg diffraction of $99.4 \%$ efficiency from a standingwave colinear to the 1D waveguide. This scheme is resilient against noisy environments, achieving 0.9~s interrogation time without any vibration noise isolation or cancellation. The AC signal sensitivity of the  scheme} can be used with the device to measure localized potentials with high sensitivity. We used this technique to measure the dynamic polarizability of the $^{39}$K ground state at 1064~nm. The interferometer may also be a useful approach to building a compact multiple-loop Sagnac atom interferometer for rotation sensing.
\end{abstract}

\maketitle

Light pulse atom interferometers are important measurement tools in areas ranging from fundamental science to practical applications such as inertial sensing~\cite{bongs2019taking,cronin2009optics}. The most widely-used atom interferometer is the free-space Mach-Zehnder configuration in which an atomic wave packet is split by a beamsplitter light pulse into two wavepackets which move apart before being reflected by a mirror light pulse and then subjected to a second beamsplitter pulse when the wavepackets overlap again. Long interrogation times are desirable because sensitivity scales as the square of the interrogation time. However, because the atoms are falling freely under gravity during the interrogation sequence the most sensitive free-space atom interferometers are several meters high~\cite{asenbaum2017phase}, which is incompatible with applications requiring a high-performance portable instrument~\cite{wu2019gravity}.
 
The guided atom interferometer, a matter wave analogy of the fiber optic interferometer, is a promising path to the goal of a sensitive compact atom interferometer~\cite{xu2019probing,zhang2016trapped,burke2016confinement,jo2007long, wang2005atom} because the waveguide supports the atoms against gravity during the interferometer sequence. Since the waveguide potential affects the propagation and phase accumulation of the matter wave packets it is essential that it be stable, flat, and smooth. While painted waveguide structures~\cite{ryu2015integrated} can be flat and smooth, it has not yet been shown that guide potential fluctuations can be made sufficiently small for atom interferometry. In waveguides formed by laser beams~\cite{mcdonald2013optically} or magnetic fields from macroscopic coils~\cite{garcia2006bose} the potential can be smooth and, if the potential is symmetric, fluctuations made common mode so that they do not affect the interferometer phase. Potentials made in this way necessarily have some curvature along the waveguide, an imperfection which can reduce the interferometer contrast over time, limiting the usable interrogation time~\cite{burke2016confinement,leonard2012effect}. The longest interrogation time reported to date for a waveguide atom interferometer appears to be 100~ms~\cite{burke2016confinement}. 

We report a guided atom interferometer exhibiting high contrast fringes with an interrogation time of over one second in a waveguide with significant curvature. The waveguide is formed by a loosely focusing 1064~nm beam, and the atom momentum is imparted by Bragg pulses formed by standing waves collinear to the waveguide. Using a non-interacting Bose-Einstein condensate as the matter wave source enables a multiple loop scheme~\cite{Jaffe2018efficient, mcguirk2002sensitive, schubert2021multi, sidorenkov2020tailoring} involving hundreds of short round trips in the guide because the contrast-degrading scattering that occurs when the wavepackets pass through each other is suppressed. Thus our results have significantly improved previous demonstration in the number of loop, contrast, and interrogation time~\cite{Jaffe2018efficient}. The fast modulation of wavepacket momentum by the high reflection rate implements a form of dynamic decoupling~\cite{hahn1950spin} or phase sensitive detection~\cite{kotler2011single} that greatly reduces the effect of technical phase noise in the interferometer, to the point where the instrument operates in a noisy environment with no vibration isolation or stabilization of the inertial reference mirror. The interferometer would be particularly well-suited to measuring rotation through the Sagnac effect~\cite{schubert2021multi, sidorenkov2020tailoring} and to probing interesting weak interactions~\cite{jaffe2017testing, haslinger2018attractive} on micron-scale distances. 
   
The experimental scheme is shown in Fig.~\ref{fig:ExperimentalScheme}. Full details are provided in the Supplemental Material~\cite{kim2022supplementary}. The interferometer sequence starts with formation of a nearly pure Bose-Einstein condensate (BEC) of $^{39}$K in the $\ket{F,m_F}=\ket{1,-1}$ state near the center of a one-dimensional waveguide with trapping frequencies $\omega_{x,y,z}=2\pi(77,100,2.8)~$Hz. The $^{39}$K isotope is useful because of a convenient Feshbach resonance~\cite{derrico2007feshbach} that allows a magnetic field to tune the s-wave scattering length through zero~\cite{fattori2008atom, roati200739K} and to control the collisional loss~\cite{band2000elastic}. The waveguide potential is a combination of the approximately harmonic transverse optical dipole potential ($\omega_x=2\pi\times77~$Hz and $\omega_y=2\pi\times100~$Hz) of a 1064~nm laser beam and a harmonic axial magnetic potential ($\omega_z=2\pi\times2.8~$Hz) arising from the residual curvature in the Helmholtz coils that control the interaction strength. An undesirable compressional mode of the BEC along the $z$-axis is induced during the condensate preparation~\cite{dicarli2019excitation, haller2009realization}. The interferometer beamsplitters and mirrors are formed by pulses of a 766~nm standing wave, aligned with the waveguide, that impart or change the wavepacket momentum by Bragg diffraction~\cite{szigeti2012why,muller2008atom,wu2005splitting}. As shown in Fig.~\ref{fig:ExperimentalScheme}(a) and (b), the first splitting pulse transforms the initially stationary condensate into a coherent superposition of two momentum states, $(\psi_++\psi_-)/\sqrt{2}$, and the subsequent series of $n$ mirror pulses each switch the direction of momentum, $\psi_+\leftrightarrow\psi_-$. The last splitting pulse with adjustable Bragg laser phase $\theta$ recombines the atoms, and the atom numbers in each momentum state ($N_{+,-,0}$) are measured by absorption imaging using a fringe suppression algorithm~\cite{niu2018optimized, ockeloen2010detection}. The interferometer phase is a linear combination of the Bragg pulse phases seen by the atoms that takes the form of a sum with alternating sign, $\phi_{laser}=\phi_{\pi/2}+[2\sum_{i=1}^{n=even} (-1)^i\phi_{n,\pi}]-\phi_{\pi/2,\theta}$~\cite{peters2001high}. 

\begin{figure}
\includegraphics[width=0.9\columnwidth]{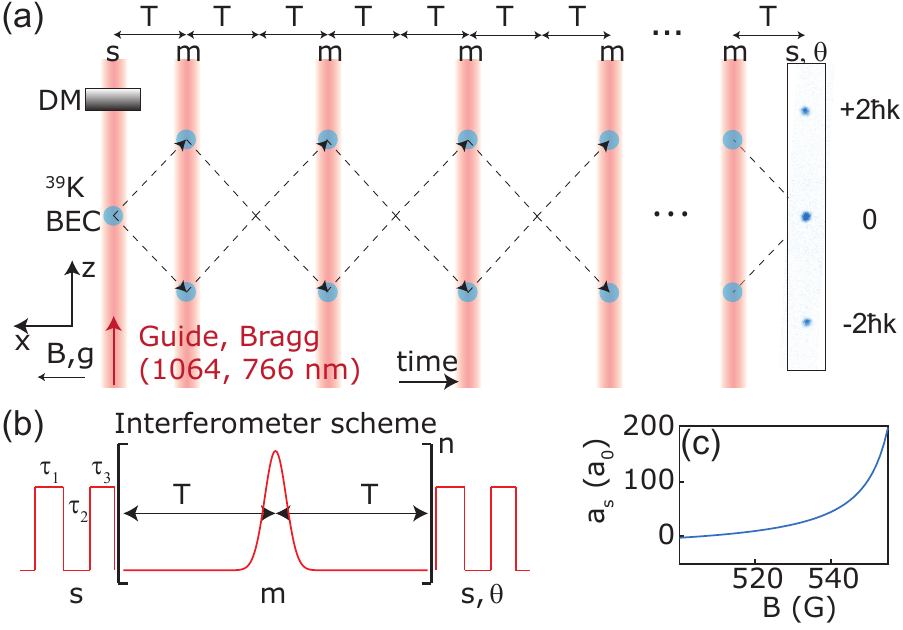}
\caption{\label{fig:ExperimentalScheme} Experimental scheme. (a) A 1064~nm laser beam creates the guide potential and a collinear 766~nm laser beam (both represented by the wide pink line) creates the optical Bragg grating by retro-reflection from a long-pass dichroic mirror (DM). A $^{39}$K $\ket{F, m_F} =\ket{1,-1}$ BEC (blue circle) is loaded into the waveguide. In the guide, two momentum states $\pm2\hbar k$, with $k=2\pi/$766~nm periodically bounce back and forth in response to a series of mirror pulses applied after the splitting. The last beamsplitter pulse recombines the wavepackets with adjustable Bragg laser phase $\theta$,  and the atom numbers are measured by absorption imaging, as shown at the right-most side. The image is a single shot, showing a $0.18~$mm$\times1.47~$mm region after 16~ms of free-expansion without the waveguide.  The black arrow labeled $B, g$ shows the direction of the Helmholtz magnetic field and gravity. (b) The interferometer scheme is composed of a series of regularly-spaced mirror pulses ($m$) sandwiched between the two beamsplitter pulses ($s$). The beamsplitter uses square pulses of duration $\tau_1$ and $\tau_3$ separated by delay $\tau_2$~\cite{wu2005splitting}, with ($\tau_1,\tau_2,\tau_3)=(9.7, 16.5, 9.2)~\mu$s and the reflection pulse is a Gaussian of $59.8~\mu$s FWHM. (c) The magnetic field dependence of the $s$-wave scattering length $a_s$, in units of Bohr radius $a_0$, around the Feshbach resonance used to tune the strength of the interatomic interaction.}
\end{figure}

\begin{figure}
\includegraphics[width=0.9\columnwidth]{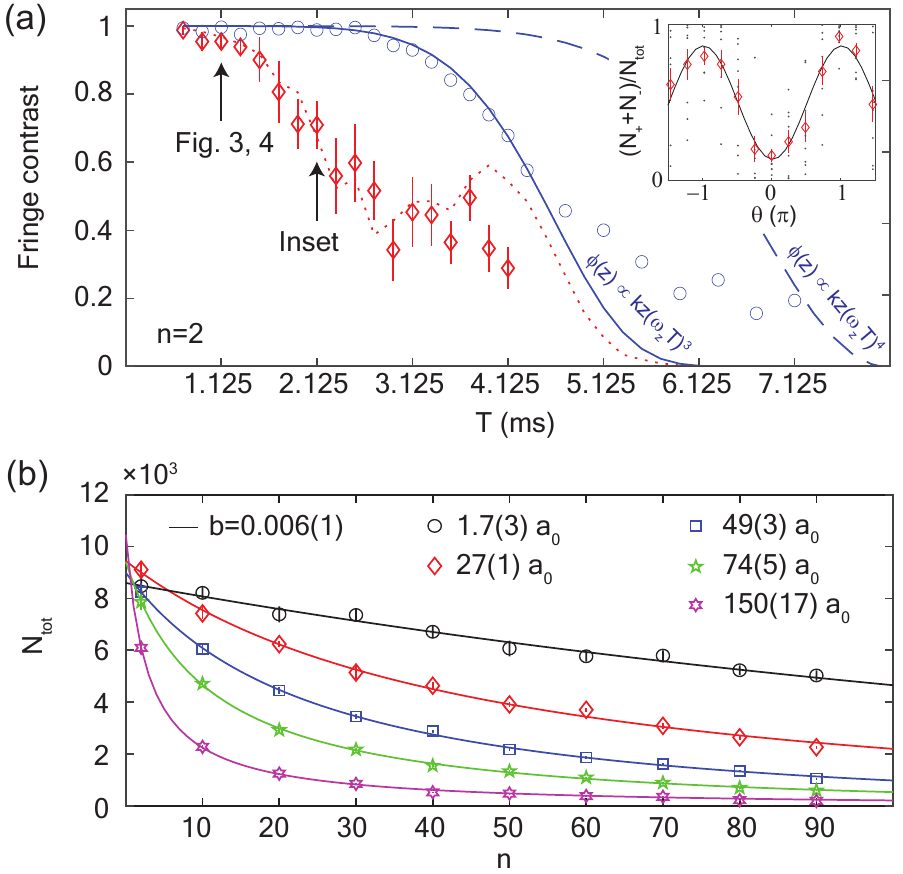} 
\caption{\label{fig:InterrogationTimeLimits} Analysis of mechanisms limiting interrogation time. (a) The fringe contrast as a function of $T$ at $n=2$ and $a_s=1.7a_0$. Blue circles and red diamonds indicate the maximum possible fringe contrast $C_{max}$ and the averaged fringe contrast $C$, respectively. Dashed and solid blue lines are the theoretical $C_{max}=\int |\psi_+\psi_-^*|\cos(\phi(z))dz$ taking into account the phase gradient from waveguide curvature, $\phi(z)\propto kz(\omega_zT)^4$, and the compressional mode, $\phi(z)\propto kz(\omega_zT)^3$, respectively. The red dotted line is a Monte-Carlo simulation of $C=C_{max}\times\overline{\cos}\delta\phi_{laser}^i$, which averages over the shot-to-shot reference mirror vibration $\delta\phi_{laser}^i$~\cite{kim2022supplementary}. (Inset) Raw data of atom split ratio. (b) Atom numbers $N_{tot}$ as a function of $n$ for different $a_s$. Symbols and lines are averaged data and numerical fits (see text). The mirror pulse efficiency is $1-b=99.4(1)\%$  (black line). $N_{tot}=N_++N_-+N_0$. Error bars are the standard error in the mean.}
\end{figure}

Figure  \ref{fig:InterrogationTimeLimits}(a) illustrates some challenges implicit in waveguide atom interferometry. The inset shows both the raw data for a fringe (gray dots, the fraction of atoms in the $\pm2\hbar k$ momentum state versus the phase of the final Bragg pulse) and the resulting averaged interference fringe (red diamonds and black line). The envelope of the raw data indicates the underlying fringe contrast $C_{max}$, which is reduced by technical phase noise to the lower value $C$ given by the amplitude of the averaged fringe. Figure \ref{fig:InterrogationTimeLimits}(a) shows both measures of contrast for a double loop ($n=2$) interferometer as a function of half the reflection pulse separation time $T$. The contrast decreases with increasing $T$ for two reasons: the axial curvature in the waveguide potential and the phase noise resulting from mechanical vibrations of the retro-reflecting mirror. The curvature effect is seen in the behavior of the maximum fringe contrast $C_{max}$ (blue circles), which shows a large reduction in contrast when $T$ exceeds a few ms due to the spatial phase gradient over the wave packet (blue lines)~\cite{burke2016confinement, kim2022supplementary}: The blue dashed line indicates the degradation of $C_{max}$ due to the spatial phase gradient that is proportional to $kz(\omega_z T)^4$, originating from that the atoms lose their momentum when climb up the waveguide while the momentum kick from the Bragg diffraction is constant~\cite{burke2016confinement}. In our particular experiment, the BEC wavepacket exhibits a compressional mode that largely fluctuates the wavepacket size over time~\cite{haller2009realization}, which further exacerbates the spatial phase gradient to $kz(\omega_z T)^3$~\cite{kim2022supplementary}. The shot-to-shot phase fluctuations from mechanical noise further reduce the average fringe contrast $C$ (red diamonds). The behavior is in good agreement with a model based on the measured mechanical motion of the reference mirror (red dotted line)~\cite{kim2022supplementary},

The main result of this paper is that using many reflections with short $T$ to confine the atoms near in the center of the guide, rather than a few large amplitude excursions in which the atoms would see more curvature, mitigates the reductions in contrast due to curvature~\cite{burke2016confinement} and to vibration-induced phase noise. One of the major concern with this approach is that in general the two wavepackets will lose atoms due to scattering every time they pass through each other. This process is illustrated in Fig.~\ref{fig:InterrogationTimeLimits}(b), where the total atom number $N_{tot}$ is plotted as a function of the number of reflections $n$ for various s-wave scattering length $a_s$. The fitted curves are the analytic solution of the differential equation $\text{d}N_{tot}/\text{d}n=-aN_{tot}^2-bN_{tot}$ describing one- and two-body loss. We can associate the coefficient $b$ with the loss rate of a mirror pulse because the exponential decay time of stationary atom numbers is longer than $20~$s for $a_s<200~a_0$. The current mirror efficiency $1-b=99.4(1)\%$ can sustain up to $n=400$ loops if the two-body loss rate is sufficiently small. We find that the two-body loss coefficient $a$ scales like $a_s^2$~\cite{kim2022supplementary}, in agreement with theory~\cite{band2000elastic}. The fitted curve at s-wave scattering length $a_s=1.7~a_0$ is indistinguishable from the case of zero two-body loss coefficient ($a=0$)~\cite{kim2022supplementary}, showing that the collisional loss can effectively be eliminated by magnetic field control of $a_s$. We used $a_s=1.7a_0$ for the rest of the experiments.
 
\begin{figure}
\includegraphics[width=1\columnwidth]{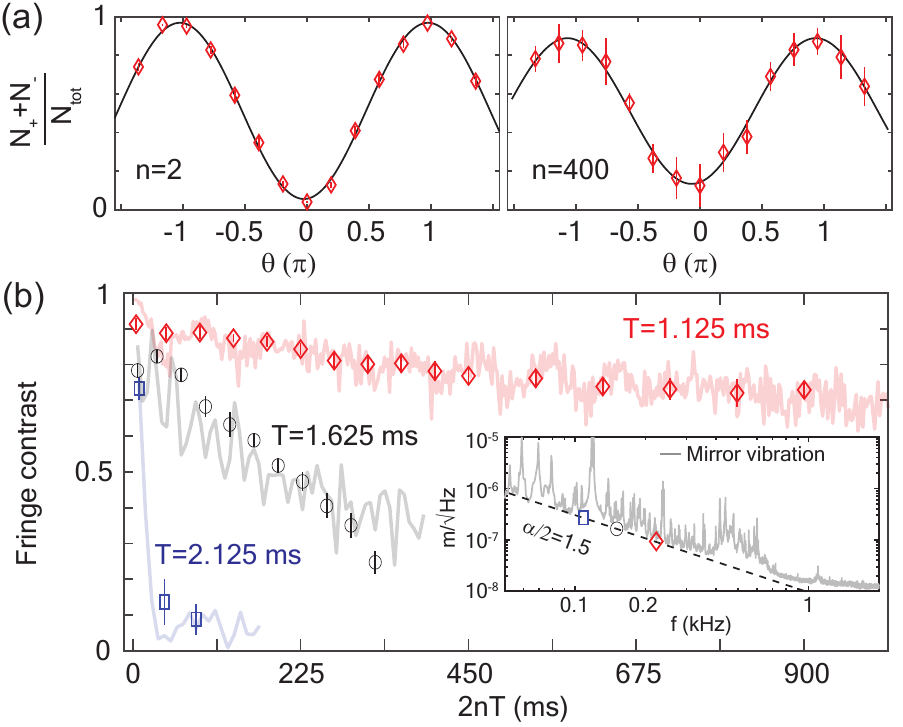} 
\caption{\label{fig:Fringes} Results of the multiple-loop atom interferometer. (a) Split ratio $(N_++N_-)/N_{tot}$ as a function of final beamsplitter phase $\theta$ for $T=1.125~$ms and $a_s=1.7a_0$. Red symbols are the averages of at least 8 shots. Black lines are fits to the function $[1-C \cos{(\theta+\phi)}]/2$, where $C=0.91(2), 0.73(3)$ and $\phi=0.07(2), 0.21(3)$ radians are respectively the fringe contrast and interferometer phase.  (b) Fringe contrast as a function of $n$ for different $T$ at $a_s=1.7a_0$. Symbols indicate the average fringe contrast. Faint solid lines are the results of a Monte-Carlo simulation~\cite{kim2022supplementary}. Inset: The noise amplitude spectrum of the reference mirror on a log-log scale. The dashed lines are the noise slope. Error bars are the standard error in the mean.}
\end{figure} 
 
Figure~\ref{fig:Fringes}(a) compares interferometer fringes corresponding to $n=2$ reflections (one round-trip) and to $n=400$ reflections (200 round-trips) for $T=1.125~$ms under this condition of negligible scattering loss. We chose $n$ to be even because the resulting symmetrical interferometer suppresses effects that would reduce fringe contrast if the number of round trips was odd~\cite{burke2016confinement, kim2022supplementary}. We have observed fringes with contrast $C=0.73(3)$ after an interrogation time $2nT=900~$ms (Fig.~\ref{fig:Fringes}(b)), which is an order of magnitude longer than previous waveguide atom interferometers~\cite{burke2016confinement,jo2007long,wang2005atom}. Moreover, the practical total interrogation time is currently limited by atom loss from the interferometer due to the slightly imperfect mirror reflectivity (black curve in Fig.~\ref{fig:InterrogationTimeLimits}(b)) because the interaction-induced dephasing is much slower than the measured interrogation times for small $a_s$ \cite{fattori2008atom,gustavsson2008control}. A number of round trips $n=2,000$ (expected interrogation time $4.5~$s) should be possible if the reflection pulse efficiency is improved to $99.9\%$ by reducing the momentum width of the atoms~\cite{szigeti2012why}. That will be possible if the amplitude of the compressional mode is reduced by employing an appropriate loading sequence of the BEC~\cite{dicarli2019excitation}. It is useful to compare our approach to that of Ref.~\cite{mcdonald2014bright} in which a bright soliton ($a_s<0$) was used to compensate for the wave packet dispersion. Here we can choose a positive $a_s$ because of the compressional mode, which will allow the atom number, and hence the signal, to be larger than is possible for the soliton approach.
   
It is remarkable that the long interrogation time is obtained without any mechanical noise cancellation or isolation-the optical table supporting the atom interferometer sits on rigid legs standing on the laboratory floor. This noise immunity arises because the form of the interferometer phase as an alternating sign sum of phases leads to an interferometer response to Bragg pulse phase noise which acts as a pass band filter of frequency $1/(4T)$, similar to lock-in detection~\cite{kotler2011single}, suggesting that the system might be useful for noise spectrum analysis~\cite{yuge2011measurement}. Figure~\ref{fig:Fringes}(b) shows that the multiple-loop configuration with short $T$ improves the available interrogation time by at least two orders of magnitude. The faint bands show the predictions of a model in which the phase of the atom-laser interaction was calculated classically as if the wave packet was a point particle~\cite{kim2022supplementary, peters2001high}. It is in good agreement with the experimental data. We observed that the fringe contrast at fixed $n$ dropped rapidly with increasing $T$ [Fig.~\ref{fig:Fringes}(b)]. We found that the phase noise scales as $\delta \phi^2_{laser}\propto 2n(2T)^\alpha$ for $n\gg1$~\cite{kim2022supplementary} and that the noise slope $\alpha$ at the region is steep enough to wash out the fringe (see inset of Fig.~\ref{fig:Fringes}(b)). This result shows that the multiple-loop scheme outperforms the single loop approach in an environment where the noise power spectral density scales like $1/f^\alpha$ when $\alpha>1$.

\begin{figure}
\includegraphics[width=0.9\columnwidth]{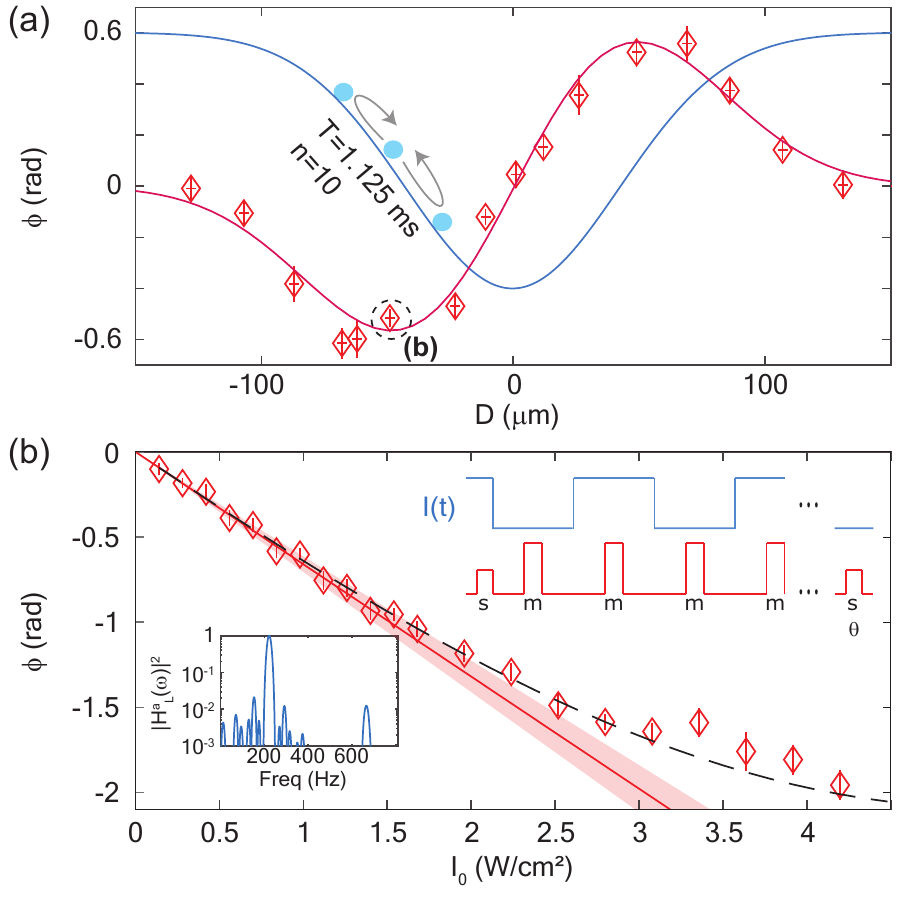} 
\caption{\label{fig:ACStark} Measurement of a high spatial gradient signal via lock-in detection. (a) Red diamonds and line are the measured and calculated interferometer phase shift respectively as a function of the distance $D$ between the center of the beam creating the AC Stark shift and the initial position of the BEC. The blue line is the 1064~nm beam profile of waist ($1/e^2$) $85(1)~\mu$m. The blue circles are schematics of how atoms maneuver in the potential (not to scale). (b) Red diamonds are the phases as a function of the peak intensity. Red line and its area are the fit of the theory curve, $U(D,t)=\alpha_{1064~nm}I(t)\exp(-2D^2/w_0^2)/(2\epsilon_0c)$, of only one free parameter ($\alpha_{1064~nm}$), and $7\%$ uncertainty of power and $w_0$ measurement, respectively. It happened that $I(t<0)=I_0$ pulls the atoms to the beam center and causes the loss of the linearity (black dashed line). Lower-left inset: The transfer function of acceleration is a band-pass filter that is sensitive to the 1064~nm intensity modulation, $I(t)=I_0[1+sign\{\sin(2\pi Ft)\}]/2$, where $F=1/(4T)$. Upper-right inset: Synchronous modulation of the potential (blue) and the interferometer pulses (red).  Error bars are the standard error in the mean.}
\end{figure}

We have used the lock-in like response of the multiple-loop atom interferometer to measure the AC Stark shift~\cite{deissler2008measurement} produced by a 1064~nm Gaussian beam of waist $w_0=85(1)~\mu$m and maximum intensity $I_0$ overlapping the waveguide and propagating perpendicular to it (Fig.~\ref{fig:ACStark}). This application highlights another advantage of the multiple-loop approach for measuring localized interactions, namely that the atoms can accumulate phase shifts from the interaction many times during the total interrogation time, thereby boosting the sensitivity. The on/off state of the AC Stark shift beam is switched every time the wavepackets cross in the center of the waveguide (upper-right inset of Fig.~\ref{fig:ACStark}(b)). This synchronous signal modulation results in the highest sensitivity as the interferometer coherently accumulates the extra phase shift from the interaction~\cite{kim2022supplementary}. The red line in Fig.~\ref{fig:ACStark}(b) is a fit of the data up to $I_0 < 1.5~W/cm^2$ (see caption). Whereas the black dashed line is a fit including the drag effect of the attractive 1064~nm potential; higher the intensity, the atoms are pulled into the intensity maximum and acumulates less phase. From this fit we find a value for the ground state polarizability of $\alpha_{1064~nm}=620(40)$ atomic units, which is in good agreement with the reference value $599.14(47)$ atomic units ~\cite{parinaz2021portal}. The multiple-loop phase-sensitive atom interferometer would be ideal for measuring interactions of the wavepackets with a localized source in which the size is comparable to or smaller than the atom separation, since the phase could accumulate coherently as the wavepackets make multiple trips past the source, becoming $n$ times the single-loop phase shift after $n$ round trips.

In addition to the possibility of phase-sensitive detection, the multiple-loop approach might offer better signal-to-noise ratio than that of a single loop in situations where mechanical vibrations transmitted to the retro-reflecting mirror create significant phase noise. An important application of the multiple-loop scheme is the case of a Sagnac atom interferometer realized by moving the waveguide in a horizontal plane transversely back and forth synchronous with the reflection pulse sequence so that the wavepacket trajectories enclose area~\cite{krzyzanowska2022matter} or by using a closed loop waveguide~\cite{ryu2015integrated, pandey2019hypersonic, moan2020quantum}. The enclosed area and hence the Sagnac phase is proportional to the number of loops, amplifying the signal without increasing the physical size of the interferometer.
 
Looking to the future, we believe that $T=20~$ms is feasible for a moderate waveguide beam waist ($100~\mu $m) with active vibration reduction and removal of waveguide curvature $\omega_z$. It will be possible to reduce $\omega_z$ by optimizing the magnetic coil geometry and reducing the field to $100~$G to exploit another Feshbach resonance where $a_s=7a_0$ and $\text{d}E/\text{d}B=34~$kHz/G for $^{39}$K $\ket{F,m_F}=\ket{1,-1}$~\cite{derrico2007feshbach}. Strontium atoms are also a good choice for atom interferometry since they have small $a_s$ and zero magnetic moment~\cite{zhang2016trapped}. The instrument that does not require large B field is also practical to use outside of the laboratory, combined with the noise-immune characteristic of the multiple-loop scheme. If the mirror pulse efficiency is optimized to $99.9\%$, an interrogation time of $60~$s should be possible. To the optimistic projection of rotation sensitivity based off of our previous demonstration of moving waveguide Sagnac interferometer in which the sensitivity was $30~\mu rad/s$ at $n=2, T=80~ms$~\cite{krzyzanowska2022matter}, our multiple-loop scheme might envision $30~nrad/s$ at $n=2,000, T=80~ms$ within a $1~mm$ scale, in which the sensitivity was only available for a meter scale baseline~\cite{savoie2018iinterleaved}.
  
We have demonstrated multiple-loop atom interferometry in a waveguide. Collisional loss is suppressed deploying a magnetic Feshbach resonance, reducing the $s$-wave scattering length close to zero. The interrogation times are extended by two orders of magnitude from the single-loop limit set by the waveguide curvature and technical noise. We have also demonstrated quantum lock-in amplification~\cite{kotler2011single}. The system may be useful for multiple-loop Sagnac interferometry~\cite{schubert2021multi}, for noise spectrum analysis~\cite{yuge2011measurement} and, perhaps, for resonant detection of gravitational waves~\cite{graham2016resonant}.

\section*{Acknowledgments}
We would like to acknowledge Saurabh Pandey for useful comments, Doga Kurkcuoglu for theoretical help, Nick Dallmann for the loan of a network analyzer, and Carlo Samson for work on the design and construction of the apparatus.
This work was supported by the U.S. Department of Energy through awards 20180045DR and 20180753PRD3 of the LANL Laboratory Directed Research and Development program.
\\ \\
$^\ast$E-mail: \textit{khs89p@gmail.com}\\
$^\ddag$E-mail: \textit{boshier@lanl.gov}\\
$^\dag$Current address: \textit{Atom Computing, Inc., Boulder, Colorado 80301, USA}

\bibliography{main.bib}

\begin{thebibliography}{52}%
\makeatletter
\providecommand \@ifxundefined [1]{%
 \@ifx{#1\undefined}
}%
\providecommand \@ifnum [1]{%
 \ifnum #1\expandafter \@firstoftwo
 \else \expandafter \@secondoftwo
 \fi
}%
\providecommand \@ifx [1]{%
 \ifx #1\expandafter \@firstoftwo
 \else \expandafter \@secondoftwo
 \fi
}%
\providecommand \natexlab [1]{#1}%
\providecommand \enquote  [1]{``#1''}%
\providecommand \bibnamefont  [1]{#1}%
\providecommand \bibfnamefont [1]{#1}%
\providecommand \citenamefont [1]{#1}%
\providecommand \href@noop [0]{\@secondoftwo}%
\providecommand \href [0]{\begingroup \@sanitize@url \@href}%
\providecommand \@href[1]{\@@startlink{#1}\@@href}%
\providecommand \@@href[1]{\endgroup#1\@@endlink}%
\providecommand \@sanitize@url [0]{\catcode `\\12\catcode `\$12\catcode
  `\&12\catcode `\#12\catcode `\^12\catcode `\_12\catcode `\%12\relax}%
\providecommand \@@startlink[1]{}%
\providecommand \@@endlink[0]{}%
\providecommand \url  [0]{\begingroup\@sanitize@url \@url }%
\providecommand \@url [1]{\endgroup\@href {#1}{\urlprefix }}%
\providecommand \urlprefix  [0]{URL }%
\providecommand \Eprint [0]{\href }%
\providecommand \doibase [0]{https://doi.org/}%
\providecommand \selectlanguage [0]{\@gobble}%
\providecommand \bibinfo  [0]{\@secondoftwo}%
\providecommand \bibfield  [0]{\@secondoftwo}%
\providecommand \translation [1]{[#1]}%
\providecommand \BibitemOpen [0]{}%
\providecommand \bibitemStop [0]{}%
\providecommand \bibitemNoStop [0]{.\EOS\space}%
\providecommand \EOS [0]{\spacefactor3000\relax}%
\providecommand \BibitemShut  [1]{\csname bibitem#1\endcsname}%
\let\auto@bib@innerbib\@empty
\bibitem [{\citenamefont {Bongs}\ \emph {et~al.}(2019)\citenamefont {Bongs},
  \citenamefont {Holynski}, \citenamefont {Vovrosh}, \citenamefont {Bouyer},
  \citenamefont {Condon}, \citenamefont {Rasel}, \citenamefont {Schubert},
  \citenamefont {Schleich},\ and\ \citenamefont {Roura}}]{bongs2019taking}%
  \BibitemOpen
  \bibfield  {author} {\bibinfo {author} {\bibfnamefont {K.}~\bibnamefont
  {Bongs}}, \bibinfo {author} {\bibfnamefont {M.}~\bibnamefont {Holynski}},
  \bibinfo {author} {\bibfnamefont {J.}~\bibnamefont {Vovrosh}}, \bibinfo
  {author} {\bibfnamefont {P.}~\bibnamefont {Bouyer}}, \bibinfo {author}
  {\bibfnamefont {G.}~\bibnamefont {Condon}}, \bibinfo {author} {\bibfnamefont
  {E.}~\bibnamefont {Rasel}}, \bibinfo {author} {\bibfnamefont
  {C.}~\bibnamefont {Schubert}}, \bibinfo {author} {\bibfnamefont {W.~P.}\
  \bibnamefont {Schleich}},\ and\ \bibinfo {author} {\bibfnamefont
  {A.}~\bibnamefont {Roura}},\ }\bibfield  {title} {\bibinfo {title} {Taking
  atom interferometric quantum sensors from the laboratory to real-world
  applications},\ }\href {https://doi.org/10.1038/s42254-019-0117-4} {\bibfield
   {journal} {\bibinfo  {journal} {Nature Reviews Physics}\ }\textbf {\bibinfo
  {volume} {1}},\ \bibinfo {pages} {731} (\bibinfo {year} {2019})}\BibitemShut
  {NoStop}%
\bibitem [{\citenamefont {Cronin}\ \emph {et~al.}(2009)\citenamefont {Cronin},
  \citenamefont {Schmiedmayer},\ and\ \citenamefont
  {Pritchard}}]{cronin2009optics}%
  \BibitemOpen
  \bibfield  {author} {\bibinfo {author} {\bibfnamefont {A.~D.}\ \bibnamefont
  {Cronin}}, \bibinfo {author} {\bibfnamefont {J.}~\bibnamefont
  {Schmiedmayer}},\ and\ \bibinfo {author} {\bibfnamefont {D.~E.}\ \bibnamefont
  {Pritchard}},\ }\bibfield  {title} {\bibinfo {title} {Optics and
  interferometry with atoms and molecules},\ }\href
  {http://dx.doi.org/10.1103/RevModPhys.81.1051} {\bibfield  {journal}
  {\bibinfo  {journal} {Reviews of Modern Physics}\ }\textbf {\bibinfo {volume}
  {81}},\ \bibinfo {pages} {1051} (\bibinfo {year} {2009})}\BibitemShut
  {NoStop}%
\bibitem [{\citenamefont {Asenbaum}\ \emph {et~al.}(2017)\citenamefont
  {Asenbaum}, \citenamefont {Overstreet}, \citenamefont {Kovachy},
  \citenamefont {Brown}, \citenamefont {Hogan},\ and\ \citenamefont
  {Kasevich}}]{asenbaum2017phase}%
  \BibitemOpen
  \bibfield  {author} {\bibinfo {author} {\bibfnamefont {P.}~\bibnamefont
  {Asenbaum}}, \bibinfo {author} {\bibfnamefont {C.}~\bibnamefont
  {Overstreet}}, \bibinfo {author} {\bibfnamefont {T.}~\bibnamefont {Kovachy}},
  \bibinfo {author} {\bibfnamefont {D.~D.}\ \bibnamefont {Brown}}, \bibinfo
  {author} {\bibfnamefont {J.~M.}\ \bibnamefont {Hogan}},\ and\ \bibinfo
  {author} {\bibfnamefont {M.~A.}\ \bibnamefont {Kasevich}},\ }\bibfield
  {title} {\bibinfo {title} {Phase shift in an atom interferometer due to
  spacetime curvature across its wave function},\ }\href
  {https://doi.org/10.1103/PhysRevLett.118.183602} {\bibfield  {journal}
  {\bibinfo  {journal} {Physical Review Letters}\ }\textbf {\bibinfo {volume}
  {118}},\ \bibinfo {pages} {183602} (\bibinfo {year} {2017})}\BibitemShut
  {NoStop}%
\bibitem [{\citenamefont {Wu}\ \emph {et~al.}(2019)\citenamefont {Wu},
  \citenamefont {Pagel}, \citenamefont {Malek}, \citenamefont {Nguyen},
  \citenamefont {Zi}, \citenamefont {Scheirer},\ and\ \citenamefont
  {M\"{u}ller}}]{wu2019gravity}%
  \BibitemOpen
  \bibfield  {author} {\bibinfo {author} {\bibfnamefont {X.}~\bibnamefont
  {Wu}}, \bibinfo {author} {\bibfnamefont {Z.}~\bibnamefont {Pagel}}, \bibinfo
  {author} {\bibfnamefont {B.~S.}\ \bibnamefont {Malek}}, \bibinfo {author}
  {\bibfnamefont {T.~H.}\ \bibnamefont {Nguyen}}, \bibinfo {author}
  {\bibfnamefont {F.}~\bibnamefont {Zi}}, \bibinfo {author} {\bibfnamefont
  {D.~S.}\ \bibnamefont {Scheirer}},\ and\ \bibinfo {author} {\bibfnamefont
  {H.}~\bibnamefont {M\"{u}ller}},\ }\bibfield  {title} {\bibinfo {title}
  {Gravity surveys using a mobile atom interferometer},\ }\href
  {https://doi.org/10.1126/sciadv.aax0800} {\bibfield  {journal} {\bibinfo
  {journal} {Science Advances}\ }\textbf {\bibinfo {volume} {5}},\ \bibinfo
  {pages} {eaax0800} (\bibinfo {year} {2019})}\BibitemShut {NoStop}%
\bibitem [{\citenamefont {Xu}\ \emph {et~al.}(2019)\citenamefont {Xu},
  \citenamefont {Jaffe}, \citenamefont {Panda}, \citenamefont {Kristensen},
  \citenamefont {Clark},\ and\ \citenamefont {M\"{u}ller}}]{xu2019probing}%
  \BibitemOpen
  \bibfield  {author} {\bibinfo {author} {\bibfnamefont {V.}~\bibnamefont
  {Xu}}, \bibinfo {author} {\bibfnamefont {M.}~\bibnamefont {Jaffe}}, \bibinfo
  {author} {\bibfnamefont {C.~D.}\ \bibnamefont {Panda}}, \bibinfo {author}
  {\bibfnamefont {S.~L.}\ \bibnamefont {Kristensen}}, \bibinfo {author}
  {\bibfnamefont {L.~W.}\ \bibnamefont {Clark}},\ and\ \bibinfo {author}
  {\bibfnamefont {H.}~\bibnamefont {M\"{u}ller}},\ }\bibfield  {title}
  {\bibinfo {title} {Probing gravity by holding atoms for 20 seconds},\ }\href
  {https://doi.org/10.1126/science.aay6428} {\bibfield  {journal} {\bibinfo
  {journal} {Science}\ }\textbf {\bibinfo {volume} {366}},\ \bibinfo {pages}
  {745} (\bibinfo {year} {2019})}\BibitemShut {NoStop}%
\bibitem [{\citenamefont {Zhang}\ \emph {et~al.}(2016)\citenamefont {Zhang},
  \citenamefont {del Aguila}, \citenamefont {Mazzoni}, \citenamefont {Poli},\
  and\ \citenamefont {Tino}}]{zhang2016trapped}%
  \BibitemOpen
  \bibfield  {author} {\bibinfo {author} {\bibfnamefont {X.}~\bibnamefont
  {Zhang}}, \bibinfo {author} {\bibfnamefont {R.~P.}\ \bibnamefont {del
  Aguila}}, \bibinfo {author} {\bibfnamefont {T.}~\bibnamefont {Mazzoni}},
  \bibinfo {author} {\bibfnamefont {N.}~\bibnamefont {Poli}},\ and\ \bibinfo
  {author} {\bibfnamefont {G.~M.}\ \bibnamefont {Tino}},\ }\bibfield  {title}
  {\bibinfo {title} {Trapped-atom interferometer with ultracold {Sr} atoms},\
  }\href {https://doi.org/10.1103/PhysRevA.94.043608} {\bibfield  {journal}
  {\bibinfo  {journal} {Physical Review A}\ }\textbf {\bibinfo {volume} {94}},\
  \bibinfo {pages} {043608} (\bibinfo {year} {2016})}\BibitemShut {NoStop}%
\bibitem [{\citenamefont {Burke}\ \emph {et~al.}(2008)\citenamefont {Burke},
  \citenamefont {Deissler}, \citenamefont {Hughes},\ and\ \citenamefont
  {Sackett}}]{burke2016confinement}%
  \BibitemOpen
  \bibfield  {author} {\bibinfo {author} {\bibfnamefont {J.~H.~T.}\
  \bibnamefont {Burke}}, \bibinfo {author} {\bibfnamefont {B.}~\bibnamefont
  {Deissler}}, \bibinfo {author} {\bibfnamefont {K.~J.}\ \bibnamefont
  {Hughes}},\ and\ \bibinfo {author} {\bibfnamefont {C.~A.}\ \bibnamefont
  {Sackett}},\ }\bibfield  {title} {\bibinfo {title} {Confinement effects in a
  guided-wave atom interferometer with millimeter-scale arm separation},\
  }\href {https://doi.org/10.1103/PhysRevA.78.023619} {\bibfield  {journal}
  {\bibinfo  {journal} {Physical Review A}\ }\textbf {\bibinfo {volume} {78}},\
  \bibinfo {pages} {023619} (\bibinfo {year} {2008})}\BibitemShut {NoStop}%
\bibitem [{\citenamefont {Jo}\ \emph {et~al.}(2007)\citenamefont {Jo},
  \citenamefont {Shin}, \citenamefont {Will}, \citenamefont {Pasquini},
  \citenamefont {Saba}, \citenamefont {Ketterle}, \citenamefont {Pritchard},
  \citenamefont {Vengalattore},\ and\ \citenamefont {Prentiss}}]{jo2007long}%
  \BibitemOpen
  \bibfield  {author} {\bibinfo {author} {\bibfnamefont {G.~B.}\ \bibnamefont
  {Jo}}, \bibinfo {author} {\bibfnamefont {Y.}~\bibnamefont {Shin}}, \bibinfo
  {author} {\bibfnamefont {S.}~\bibnamefont {Will}}, \bibinfo {author}
  {\bibfnamefont {T.~A.}\ \bibnamefont {Pasquini}}, \bibinfo {author}
  {\bibfnamefont {M.}~\bibnamefont {Saba}}, \bibinfo {author} {\bibfnamefont
  {W.}~\bibnamefont {Ketterle}}, \bibinfo {author} {\bibfnamefont {D.~E.}\
  \bibnamefont {Pritchard}}, \bibinfo {author} {\bibfnamefont {M.}~\bibnamefont
  {Vengalattore}},\ and\ \bibinfo {author} {\bibfnamefont {M.}~\bibnamefont
  {Prentiss}},\ }\bibfield  {title} {\bibinfo {title} {Long phase coherence
  time and number squeezing of two {Bose-Einstein} condensates on an atom
  chip},\ }\href {http://link.aps.org/doi/10.1103/PhysRevLett.98.030407}
  {\bibfield  {journal} {\bibinfo  {journal} {Physical Review Letters}\
  }\textbf {\bibinfo {volume} {98}},\ \bibinfo {pages} {030407} (\bibinfo
  {year} {2007})}\BibitemShut {NoStop}%
\bibitem [{\citenamefont {Wang}\ \emph {et~al.}(2005)\citenamefont {Wang},
  \citenamefont {Anderson}, \citenamefont {Bright}, \citenamefont {Cornell},
  \citenamefont {Diot}, \citenamefont {Kishimoto}, \citenamefont {Prentiss},
  \citenamefont {Saravanan}, \citenamefont {Segal},\ and\ \citenamefont
  {Wu}}]{wang2005atom}%
  \BibitemOpen
  \bibfield  {author} {\bibinfo {author} {\bibfnamefont {Y.-J.}\ \bibnamefont
  {Wang}}, \bibinfo {author} {\bibfnamefont {D.~Z.}\ \bibnamefont {Anderson}},
  \bibinfo {author} {\bibfnamefont {V.~M.}\ \bibnamefont {Bright}}, \bibinfo
  {author} {\bibfnamefont {E.~A.}\ \bibnamefont {Cornell}}, \bibinfo {author}
  {\bibfnamefont {Q.}~\bibnamefont {Diot}}, \bibinfo {author} {\bibfnamefont
  {T.}~\bibnamefont {Kishimoto}}, \bibinfo {author} {\bibfnamefont
  {M.}~\bibnamefont {Prentiss}}, \bibinfo {author} {\bibfnamefont {R.~A.}\
  \bibnamefont {Saravanan}}, \bibinfo {author} {\bibfnamefont {S.~R.}\
  \bibnamefont {Segal}},\ and\ \bibinfo {author} {\bibfnamefont
  {S.}~\bibnamefont {Wu}},\ }\bibfield  {title} {\bibinfo {title} {Atom
  {Michelson} interferometer on a chip using a {Bose-Einstein} condensate},\
  }\href {http://link.aps.org/doi/10.1103/PhysRevLett.94.090405} {\bibfield
  {journal} {\bibinfo  {journal} {Physical Review Letters}\ }\textbf {\bibinfo
  {volume} {94}},\ \bibinfo {pages} {090405} (\bibinfo {year}
  {2005})}\BibitemShut {NoStop}%
\bibitem [{\citenamefont {Ryu}\ and\ \citenamefont
  {Boshier}(2015)}]{ryu2015integrated}%
  \BibitemOpen
  \bibfield  {author} {\bibinfo {author} {\bibfnamefont {C.}~\bibnamefont
  {Ryu}}\ and\ \bibinfo {author} {\bibfnamefont {M.~G.}\ \bibnamefont
  {Boshier}},\ }\bibfield  {title} {\bibinfo {title} {Integrated coherent
  matter wave circuits},\ }\href {http://doi.org/10.1088/1367-2630/17/9/092002}
  {\bibfield  {journal} {\bibinfo  {journal} {New Journal of Physics}\ }\textbf
  {\bibinfo {volume} {17}},\ \bibinfo {pages} {092002} (\bibinfo {year}
  {2015})}\BibitemShut {NoStop}%
\bibitem [{\citenamefont {McDonald}\ \emph {et~al.}(2013)\citenamefont
  {McDonald}, \citenamefont {Keal}, \citenamefont {Altin}, \citenamefont
  {Debs}, \citenamefont {Bennetts}, \citenamefont {Kuhn}, \citenamefont
  {Hardman}, \citenamefont {Johnsson}, \citenamefont {Close},\ and\
  \citenamefont {Robins}}]{mcdonald2013optically}%
  \BibitemOpen
  \bibfield  {author} {\bibinfo {author} {\bibfnamefont {G.~D.}\ \bibnamefont
  {McDonald}}, \bibinfo {author} {\bibfnamefont {H.}~\bibnamefont {Keal}},
  \bibinfo {author} {\bibfnamefont {P.~A.}\ \bibnamefont {Altin}}, \bibinfo
  {author} {\bibfnamefont {J.~E.}\ \bibnamefont {Debs}}, \bibinfo {author}
  {\bibfnamefont {S.}~\bibnamefont {Bennetts}}, \bibinfo {author}
  {\bibfnamefont {C.~C.~N.}\ \bibnamefont {Kuhn}}, \bibinfo {author}
  {\bibfnamefont {K.~S.}\ \bibnamefont {Hardman}}, \bibinfo {author}
  {\bibfnamefont {M.~T.}\ \bibnamefont {Johnsson}}, \bibinfo {author}
  {\bibfnamefont {J.~D.}\ \bibnamefont {Close}},\ and\ \bibinfo {author}
  {\bibfnamefont {N.~P.}\ \bibnamefont {Robins}},\ }\bibfield  {title}
  {\bibinfo {title} {Optically guided linear {Mach-Zehnder} atom
  interferometer},\ }\href {http://link.aps.org/doi/10.1103/PhysRevA.87.013632}
  {\bibfield  {journal} {\bibinfo  {journal} {Physical Review A}\ }\textbf
  {\bibinfo {volume} {87}},\ \bibinfo {pages} {013632} (\bibinfo {year}
  {2013})}\BibitemShut {NoStop}%
\bibitem [{\citenamefont {Garcia}\ \emph {et~al.}(2006)\citenamefont {Garcia},
  \citenamefont {Deissler}, \citenamefont {Hughes}, \citenamefont {Reeves},\
  and\ \citenamefont {Sackett}}]{garcia2006bose}%
  \BibitemOpen
  \bibfield  {author} {\bibinfo {author} {\bibfnamefont {O.}~\bibnamefont
  {Garcia}}, \bibinfo {author} {\bibfnamefont {B.}~\bibnamefont {Deissler}},
  \bibinfo {author} {\bibfnamefont {K.~J.}\ \bibnamefont {Hughes}}, \bibinfo
  {author} {\bibfnamefont {J.~M.}\ \bibnamefont {Reeves}},\ and\ \bibinfo
  {author} {\bibfnamefont {C.~A.}\ \bibnamefont {Sackett}},\ }\bibfield
  {title} {\bibinfo {title} {{Bose-Einstein}-condensate interferometer with
  macroscopic arm separation},\ }\href
  {http://link.aps.org/doi/10.1103/PhysRevA.74.031601} {\bibfield  {journal}
  {\bibinfo  {journal} {Physical Review A}\ }\textbf {\bibinfo {volume} {74}},\
  \bibinfo {pages} {031601} (\bibinfo {year} {2006})}\BibitemShut {NoStop}%
\bibitem [{\citenamefont {Leonard}\ and\ \citenamefont
  {Sackett}(2012)}]{leonard2012effect}%
  \BibitemOpen
  \bibfield  {author} {\bibinfo {author} {\bibfnamefont {R.~H.}\ \bibnamefont
  {Leonard}}\ and\ \bibinfo {author} {\bibfnamefont {C.~A.}\ \bibnamefont
  {Sackett}},\ }\bibfield  {title} {\bibinfo {title} {Effect of trap
  anharmonicity on a free-oscillation atom interferometer},\ }\href
  {https://doi.org/10.1103/PhysRevA.86.043613} {\bibfield  {journal} {\bibinfo
  {journal} {Physical Review A}\ }\textbf {\bibinfo {volume} {86}},\ \bibinfo
  {pages} {043613} (\bibinfo {year} {2012})}\BibitemShut {NoStop}%
\bibitem [{\citenamefont {Jaffe}\ \emph {et~al.}(2018)\citenamefont {Jaffe},
  \citenamefont {Xu}, \citenamefont {Haslinger}, \citenamefont {Müller},\ and\
  \citenamefont {Hamilton}}]{Jaffe2018efficient}%
  \BibitemOpen
  \bibfield  {author} {\bibinfo {author} {\bibfnamefont {M.}~\bibnamefont
  {Jaffe}}, \bibinfo {author} {\bibfnamefont {V.}~\bibnamefont {Xu}}, \bibinfo
  {author} {\bibfnamefont {P.}~\bibnamefont {Haslinger}}, \bibinfo {author}
  {\bibfnamefont {H.}~\bibnamefont {Müller}},\ and\ \bibinfo {author}
  {\bibfnamefont {P.}~\bibnamefont {Hamilton}},\ }\bibfield  {title} {\bibinfo
  {title} {Efficient adiabatic spin-dependent kicks in an atom
  interferometer},\ }\href {https://doi.org/10.1103/PhysRevLett.121.040402}
  {\bibfield  {journal} {\bibinfo  {journal} {Physical Review Letters}\
  }\textbf {\bibinfo {volume} {121}} (\bibinfo {year} {2018})}\BibitemShut
  {NoStop}%
\bibitem [{\citenamefont {McGuirk}\ \emph {et~al.}(2002)\citenamefont
  {McGuirk}, \citenamefont {Foster}, \citenamefont {Fixler}, \citenamefont
  {Snadden},\ and\ \citenamefont {Kasevich}}]{mcguirk2002sensitive}%
  \BibitemOpen
  \bibfield  {author} {\bibinfo {author} {\bibfnamefont {J.~M.}\ \bibnamefont
  {McGuirk}}, \bibinfo {author} {\bibfnamefont {G.~T.}\ \bibnamefont {Foster}},
  \bibinfo {author} {\bibfnamefont {J.~B.}\ \bibnamefont {Fixler}}, \bibinfo
  {author} {\bibfnamefont {M.~J.}\ \bibnamefont {Snadden}},\ and\ \bibinfo
  {author} {\bibfnamefont {M.~A.}\ \bibnamefont {Kasevich}},\ }\bibfield
  {title} {\bibinfo {title} {Sensitive absolute-gravity gradiometry using atom
  interferometry},\ }\href {https://doi.org/10.1103/PhysRevA.65.033608}
  {\bibfield  {journal} {\bibinfo  {journal} {Physical Review A}\ }\textbf
  {\bibinfo {volume} {65}},\ \bibinfo {pages} {033608} (\bibinfo {year}
  {2002})}\BibitemShut {NoStop}%
\bibitem [{\citenamefont {Schubert}\ \emph {et~al.}(2021)\citenamefont
  {Schubert}, \citenamefont {Abend}, \citenamefont {Gersemann}, \citenamefont
  {Gebbe}, \citenamefont {Schlippert}, \citenamefont {Berg},\ and\
  \citenamefont {Rasel}}]{schubert2021multi}%
  \BibitemOpen
  \bibfield  {author} {\bibinfo {author} {\bibfnamefont {C.}~\bibnamefont
  {Schubert}}, \bibinfo {author} {\bibfnamefont {S.}~\bibnamefont {Abend}},
  \bibinfo {author} {\bibfnamefont {M.}~\bibnamefont {Gersemann}}, \bibinfo
  {author} {\bibfnamefont {M.}~\bibnamefont {Gebbe}}, \bibinfo {author}
  {\bibfnamefont {D.}~\bibnamefont {Schlippert}}, \bibinfo {author}
  {\bibfnamefont {P.}~\bibnamefont {Berg}},\ and\ \bibinfo {author}
  {\bibfnamefont {E.~M.}\ \bibnamefont {Rasel}},\ }\bibfield  {title} {\bibinfo
  {title} {Multi-loop atomic {Sagnac} interferometry},\ }\href
  {https://doi.org/10.1038/s41598-021-95334-7} {\bibfield  {journal} {\bibinfo
  {journal} {Scientific Reports}\ }\textbf {\bibinfo {volume} {11}},\ \bibinfo
  {pages} {16121} (\bibinfo {year} {2021})}\BibitemShut {NoStop}%
\bibitem [{\citenamefont {Sidorenkov}\ \emph {et~al.}(2020)\citenamefont
  {Sidorenkov}, \citenamefont {Gautier}, \citenamefont {Altorio}, \citenamefont
  {Geiger},\ and\ \citenamefont {Landragin}}]{sidorenkov2020tailoring}%
  \BibitemOpen
  \bibfield  {author} {\bibinfo {author} {\bibfnamefont {L.~A.}\ \bibnamefont
  {Sidorenkov}}, \bibinfo {author} {\bibfnamefont {R.}~\bibnamefont {Gautier}},
  \bibinfo {author} {\bibfnamefont {M.}~\bibnamefont {Altorio}}, \bibinfo
  {author} {\bibfnamefont {R.}~\bibnamefont {Geiger}},\ and\ \bibinfo {author}
  {\bibfnamefont {A.}~\bibnamefont {Landragin}},\ }\bibfield  {title} {\bibinfo
  {title} {Tailoring multiloop atom interferometers with adjustable momentum
  transfer},\ }\href {https://doi.org/10.1103/PhysRevLett.125.213201}
  {\bibfield  {journal} {\bibinfo  {journal} {Physical Review Letters}\
  }\textbf {\bibinfo {volume} {125}},\ \bibinfo {pages} {213201} (\bibinfo
  {year} {2020})}\BibitemShut {NoStop}%
\bibitem [{\citenamefont {Hahn}(1950)}]{hahn1950spin}%
  \BibitemOpen
  \bibfield  {author} {\bibinfo {author} {\bibfnamefont {E.~L.}\ \bibnamefont
  {Hahn}},\ }\bibfield  {title} {\bibinfo {title} {Spin echoes},\ }\href
  {https://doi.org/10.1103/PhysRev.80.580} {\bibfield  {journal} {\bibinfo
  {journal} {Physical Review}\ }\textbf {\bibinfo {volume} {80}},\ \bibinfo
  {pages} {580} (\bibinfo {year} {1950})}\BibitemShut {NoStop}%
\bibitem [{\citenamefont {Kotler}\ \emph {et~al.}(2011)\citenamefont {Kotler},
  \citenamefont {Akerman}, \citenamefont {Glickman}, \citenamefont {Keselman},\
  and\ \citenamefont {Ozeri}}]{kotler2011single}%
  \BibitemOpen
  \bibfield  {author} {\bibinfo {author} {\bibfnamefont {S.}~\bibnamefont
  {Kotler}}, \bibinfo {author} {\bibfnamefont {N.}~\bibnamefont {Akerman}},
  \bibinfo {author} {\bibfnamefont {Y.}~\bibnamefont {Glickman}}, \bibinfo
  {author} {\bibfnamefont {A.}~\bibnamefont {Keselman}},\ and\ \bibinfo
  {author} {\bibfnamefont {R.}~\bibnamefont {Ozeri}},\ }\bibfield  {title}
  {\bibinfo {title} {Single-ion quantum lock-in amplifier},\ }\href
  {https://doi.org/10.1038/nature10010} {\bibfield  {journal} {\bibinfo
  {journal} {Nature}\ }\textbf {\bibinfo {volume} {473}},\ \bibinfo {pages}
  {61} (\bibinfo {year} {2011})}\BibitemShut {NoStop}%
\bibitem [{\citenamefont {Jaffe}\ \emph {et~al.}(2017)\citenamefont {Jaffe},
  \citenamefont {Haslinger}, \citenamefont {Xu}, \citenamefont {Hamilton},
  \citenamefont {Upadhye}, \citenamefont {Elder}, \citenamefont {Khoury},\ and\
  \citenamefont {M\"{u}ller}}]{jaffe2017testing}%
  \BibitemOpen
  \bibfield  {author} {\bibinfo {author} {\bibfnamefont {M.}~\bibnamefont
  {Jaffe}}, \bibinfo {author} {\bibfnamefont {P.}~\bibnamefont {Haslinger}},
  \bibinfo {author} {\bibfnamefont {V.}~\bibnamefont {Xu}}, \bibinfo {author}
  {\bibfnamefont {P.}~\bibnamefont {Hamilton}}, \bibinfo {author}
  {\bibfnamefont {A.}~\bibnamefont {Upadhye}}, \bibinfo {author} {\bibfnamefont
  {B.}~\bibnamefont {Elder}}, \bibinfo {author} {\bibfnamefont
  {J.}~\bibnamefont {Khoury}},\ and\ \bibinfo {author} {\bibfnamefont
  {H.}~\bibnamefont {M\"{u}ller}},\ }\bibfield  {title} {\bibinfo {title}
  {Testing sub-gravitational forces on atoms from a miniature in-vacuum source
  mass},\ }\href {https://doi.org/10.1038/nphys4189} {\bibfield  {journal}
  {\bibinfo  {journal} {Nature Physics}\ }\textbf {\bibinfo {volume} {13}},\
  \bibinfo {pages} {938} (\bibinfo {year} {2017})}\BibitemShut {NoStop}%
\bibitem [{\citenamefont {Haslinger}\ \emph {et~al.}(2018)\citenamefont
  {Haslinger}, \citenamefont {Jaffe}, \citenamefont {Xu}, \citenamefont
  {Schwartz}, \citenamefont {Sonnleitner}, \citenamefont {Ritsch-Marte},
  \citenamefont {Ritsch},\ and\ \citenamefont
  {M\"{u}ller}}]{haslinger2018attractive}%
  \BibitemOpen
  \bibfield  {author} {\bibinfo {author} {\bibfnamefont {P.}~\bibnamefont
  {Haslinger}}, \bibinfo {author} {\bibfnamefont {M.}~\bibnamefont {Jaffe}},
  \bibinfo {author} {\bibfnamefont {V.}~\bibnamefont {Xu}}, \bibinfo {author}
  {\bibfnamefont {O.}~\bibnamefont {Schwartz}}, \bibinfo {author}
  {\bibfnamefont {M.}~\bibnamefont {Sonnleitner}}, \bibinfo {author}
  {\bibfnamefont {M.}~\bibnamefont {Ritsch-Marte}}, \bibinfo {author}
  {\bibfnamefont {H.}~\bibnamefont {Ritsch}},\ and\ \bibinfo {author}
  {\bibfnamefont {H.}~\bibnamefont {M\"{u}ller}},\ }\bibfield  {title}
  {\bibinfo {title} {Attractive force on atoms due to blackbody radiation},\
  }\href {https://doi.org/10.1038/s41567-017-0004-9} {\bibfield  {journal}
  {\bibinfo  {journal} {Nature Physics}\ }\textbf {\bibinfo {volume} {14}},\
  \bibinfo {pages} {257} (\bibinfo {year} {2018})}\BibitemShut {NoStop}%
\bibitem [{kim()}]{kim2022supplementary}%
  \BibitemOpen
  \href@noop {} {}\bibinfo {note} {The Supplemental Material provides details
  of the experimental system and the interferometer sequence, along with a
  discussion of mechanisms that cause phase noise and that reduce the fringe
  contrast. It also presents the data and analysis for the study of cold
  collisions, and it provides additional information on the quantum lock-in
  amplification demonstration.}\BibitemShut {Stop}%
\bibitem [{\citenamefont {D'Errico}\ \emph {et~al.}(2007)\citenamefont
  {D'Errico}, \citenamefont {Zaccanti}, \citenamefont {Fattori}, \citenamefont
  {Roati}, \citenamefont {Inguscio}, \citenamefont {Modugno},\ and\
  \citenamefont {Simoni}}]{derrico2007feshbach}%
  \BibitemOpen
  \bibfield  {author} {\bibinfo {author} {\bibfnamefont {C.}~\bibnamefont
  {D'Errico}}, \bibinfo {author} {\bibfnamefont {M.}~\bibnamefont {Zaccanti}},
  \bibinfo {author} {\bibfnamefont {M.}~\bibnamefont {Fattori}}, \bibinfo
  {author} {\bibfnamefont {G.}~\bibnamefont {Roati}}, \bibinfo {author}
  {\bibfnamefont {M.}~\bibnamefont {Inguscio}}, \bibinfo {author}
  {\bibfnamefont {G.}~\bibnamefont {Modugno}},\ and\ \bibinfo {author}
  {\bibfnamefont {A.}~\bibnamefont {Simoni}},\ }\bibfield  {title} {\bibinfo
  {title} {{Feshbach} resonances in ultracold $^{39}\text{K}$},\ }\href
  {https://doi.org/10.1088/1367-2630/9/7/223} {\bibfield  {journal} {\bibinfo
  {journal} {New Journal of Physics}\ }\textbf {\bibinfo {volume} {9}},\
  \bibinfo {pages} {223} (\bibinfo {year} {2007})}\BibitemShut {NoStop}%
\bibitem [{\citenamefont {Fattori}\ \emph {et~al.}(2008)\citenamefont
  {Fattori}, \citenamefont {D’Errico}, \citenamefont {Roati}, \citenamefont
  {Zaccanti}, \citenamefont {Jona-Lasinio}, \citenamefont {Modugno},
  \citenamefont {Inguscio},\ and\ \citenamefont {Modugno}}]{fattori2008atom}%
  \BibitemOpen
  \bibfield  {author} {\bibinfo {author} {\bibfnamefont {M.}~\bibnamefont
  {Fattori}}, \bibinfo {author} {\bibfnamefont {C.}~\bibnamefont {D’Errico}},
  \bibinfo {author} {\bibfnamefont {G.}~\bibnamefont {Roati}}, \bibinfo
  {author} {\bibfnamefont {M.}~\bibnamefont {Zaccanti}}, \bibinfo {author}
  {\bibfnamefont {M.}~\bibnamefont {Jona-Lasinio}}, \bibinfo {author}
  {\bibfnamefont {M.}~\bibnamefont {Modugno}}, \bibinfo {author} {\bibfnamefont
  {M.}~\bibnamefont {Inguscio}},\ and\ \bibinfo {author} {\bibfnamefont
  {G.}~\bibnamefont {Modugno}},\ }\bibfield  {title} {\bibinfo {title} {Atom
  interferometry with a weakly interacting {Bose-Einstein} condensate},\ }\href
  {http://dx.doi.org/10.1103/PhysRevLett.100.080405} {\bibfield  {journal}
  {\bibinfo  {journal} {Physical Review Letters}\ }\textbf {\bibinfo {volume}
  {100}},\ \bibinfo {pages} {080405} (\bibinfo {year} {2008})}\BibitemShut
  {NoStop}%
\bibitem [{\citenamefont {Roati}\ \emph {et~al.}(2007)\citenamefont {Roati},
  \citenamefont {Zaccanti}, \citenamefont {D’Errico}, \citenamefont {Catani},
  \citenamefont {Modugno}, \citenamefont {Simoni}, \citenamefont {Inguscio},\
  and\ \citenamefont {Modugno}}]{roati200739K}%
  \BibitemOpen
  \bibfield  {author} {\bibinfo {author} {\bibfnamefont {G.}~\bibnamefont
  {Roati}}, \bibinfo {author} {\bibfnamefont {M.}~\bibnamefont {Zaccanti}},
  \bibinfo {author} {\bibfnamefont {C.}~\bibnamefont {D’Errico}}, \bibinfo
  {author} {\bibfnamefont {J.}~\bibnamefont {Catani}}, \bibinfo {author}
  {\bibfnamefont {M.}~\bibnamefont {Modugno}}, \bibinfo {author} {\bibfnamefont
  {A.}~\bibnamefont {Simoni}}, \bibinfo {author} {\bibfnamefont
  {M.}~\bibnamefont {Inguscio}},\ and\ \bibinfo {author} {\bibfnamefont
  {G.}~\bibnamefont {Modugno}},\ }\bibfield  {title} {\bibinfo {title}
  {$^{39}\text{K}$ {Bose-Einstein} condensate with tunable interactions},\
  }\href {https://doi.org/10.1103/PhysRevLett.99.010403} {\bibfield  {journal}
  {\bibinfo  {journal} {Physical Review Letters}\ }\textbf {\bibinfo {volume}
  {99}},\ \bibinfo {pages} {010403} (\bibinfo {year} {2007})}\BibitemShut
  {NoStop}%
\bibitem [{\citenamefont {Band}\ \emph {et~al.}(2000)\citenamefont {Band},
  \citenamefont {Trippenbach}, \citenamefont {Burke},\ and\ \citenamefont
  {Julienne}}]{band2000elastic}%
  \BibitemOpen
  \bibfield  {author} {\bibinfo {author} {\bibfnamefont {Y.~B.}\ \bibnamefont
  {Band}}, \bibinfo {author} {\bibfnamefont {M.}~\bibnamefont {Trippenbach}},
  \bibinfo {author} {\bibfnamefont {J.~P.}\ \bibnamefont {Burke}},\ and\
  \bibinfo {author} {\bibfnamefont {P.~S.}\ \bibnamefont {Julienne}},\
  }\bibfield  {title} {\bibinfo {title} {Elastic scattering loss of atoms from
  colliding {Bose-Einstein} condensate wave packets},\ }\href
  {https://doi.org/10.1103/PhysRevLett.84.5462} {\bibfield  {journal} {\bibinfo
   {journal} {Physical Review Letters}\ }\textbf {\bibinfo {volume} {84}},\
  \bibinfo {pages} {5462} (\bibinfo {year} {2000})}\BibitemShut {NoStop}%
\bibitem [{\citenamefont {Di~Carli}\ \emph {et~al.}(2019)\citenamefont
  {Di~Carli}, \citenamefont {Colquhoun}, \citenamefont {Henderson},
  \citenamefont {Flannigan}, \citenamefont {Oppo}, \citenamefont {Daley},
  \citenamefont {Kuhr},\ and\ \citenamefont {Haller}}]{dicarli2019excitation}%
  \BibitemOpen
  \bibfield  {author} {\bibinfo {author} {\bibfnamefont {A.}~\bibnamefont
  {Di~Carli}}, \bibinfo {author} {\bibfnamefont {C.~D.}\ \bibnamefont
  {Colquhoun}}, \bibinfo {author} {\bibfnamefont {G.}~\bibnamefont
  {Henderson}}, \bibinfo {author} {\bibfnamefont {S.}~\bibnamefont
  {Flannigan}}, \bibinfo {author} {\bibfnamefont {G.-L.}\ \bibnamefont {Oppo}},
  \bibinfo {author} {\bibfnamefont {A.~J.}\ \bibnamefont {Daley}}, \bibinfo
  {author} {\bibfnamefont {S.}~\bibnamefont {Kuhr}},\ and\ \bibinfo {author}
  {\bibfnamefont {E.}~\bibnamefont {Haller}},\ }\bibfield  {title} {\bibinfo
  {title} {Excitation modes of bright matter-wave solitons},\ }\href
  {https://doi.org/10.1103/PhysRevLett.123.123602} {\bibfield  {journal}
  {\bibinfo  {journal} {Physical Review Letters}\ }\textbf {\bibinfo {volume}
  {123}},\ \bibinfo {pages} {123602} (\bibinfo {year} {2019})}\BibitemShut
  {NoStop}%
\bibitem [{\citenamefont {Haller}\ \emph {et~al.}(2009)\citenamefont {Haller},
  \citenamefont {Gustavsson}, \citenamefont {Mark}, \citenamefont {Danzl},
  \citenamefont {Hart}, \citenamefont {Pupillo},\ and\ \citenamefont
  {Nägerl}}]{haller2009realization}%
  \BibitemOpen
  \bibfield  {author} {\bibinfo {author} {\bibfnamefont {E.}~\bibnamefont
  {Haller}}, \bibinfo {author} {\bibfnamefont {M.}~\bibnamefont {Gustavsson}},
  \bibinfo {author} {\bibfnamefont {M.~J.}\ \bibnamefont {Mark}}, \bibinfo
  {author} {\bibfnamefont {J.~G.}\ \bibnamefont {Danzl}}, \bibinfo {author}
  {\bibfnamefont {R.}~\bibnamefont {Hart}}, \bibinfo {author} {\bibfnamefont
  {G.}~\bibnamefont {Pupillo}},\ and\ \bibinfo {author} {\bibfnamefont {H.-C.}\
  \bibnamefont {Nägerl}},\ }\bibfield  {title} {\bibinfo {title} {Realization
  of an excited, strongly correlated quantum gas phase},\ }\href
  {https://doi.org/10.1126/science.1175850} {\bibfield  {journal} {\bibinfo
  {journal} {Science}\ }\textbf {\bibinfo {volume} {325}},\ \bibinfo {pages}
  {1224} (\bibinfo {year} {2009})}\BibitemShut {NoStop}%
\bibitem [{\citenamefont {Szigeti}\ \emph {et~al.}(2012)\citenamefont
  {Szigeti}, \citenamefont {Debs}, \citenamefont {Hope}, \citenamefont
  {Robins},\ and\ \citenamefont {Close}}]{szigeti2012why}%
  \BibitemOpen
  \bibfield  {author} {\bibinfo {author} {\bibfnamefont {S.~S.}\ \bibnamefont
  {Szigeti}}, \bibinfo {author} {\bibfnamefont {J.~E.}\ \bibnamefont {Debs}},
  \bibinfo {author} {\bibfnamefont {J.~J.}\ \bibnamefont {Hope}}, \bibinfo
  {author} {\bibfnamefont {N.~P.}\ \bibnamefont {Robins}},\ and\ \bibinfo
  {author} {\bibfnamefont {J.~D.}\ \bibnamefont {Close}},\ }\bibfield  {title}
  {\bibinfo {title} {Why momentum width matters for atom interferometry with
  {Bragg} pulses},\ }\href {https://doi.org/10.1088/1367-2630/14/2/023009}
  {\bibfield  {journal} {\bibinfo  {journal} {New Journal of Physics}\ }\textbf
  {\bibinfo {volume} {14}},\ \bibinfo {pages} {023009} (\bibinfo {year}
  {2012})}\BibitemShut {NoStop}%
\bibitem [{\citenamefont {M\"{u}ller}\ \emph {et~al.}(2008)\citenamefont
  {M\"{u}ller}, \citenamefont {Chiow},\ and\ \citenamefont
  {Chu}}]{muller2008atom}%
  \BibitemOpen
  \bibfield  {author} {\bibinfo {author} {\bibfnamefont {H.}~\bibnamefont
  {M\"{u}ller}}, \bibinfo {author} {\bibfnamefont {S.-w.}\ \bibnamefont
  {Chiow}},\ and\ \bibinfo {author} {\bibfnamefont {S.}~\bibnamefont {Chu}},\
  }\bibfield  {title} {\bibinfo {title} {Atom-wave diffraction between the
  raman-nath and the {Bragg} regime: Effective {Rabi} frequency, losses, and
  phase shifts},\ }\href {https://doi.org/10.1103/PhysRevA.77.023609}
  {\bibfield  {journal} {\bibinfo  {journal} {Physical Review A}\ }\textbf
  {\bibinfo {volume} {77}},\ \bibinfo {pages} {023609} (\bibinfo {year}
  {2008})}\BibitemShut {NoStop}%
\bibitem [{\citenamefont {Wu}\ \emph {et~al.}(2005)\citenamefont {Wu},
  \citenamefont {Wang}, \citenamefont {Diot},\ and\ \citenamefont
  {Prentiss}}]{wu2005splitting}%
  \BibitemOpen
  \bibfield  {author} {\bibinfo {author} {\bibfnamefont {S.}~\bibnamefont
  {Wu}}, \bibinfo {author} {\bibfnamefont {Y.-J.}\ \bibnamefont {Wang}},
  \bibinfo {author} {\bibfnamefont {Q.}~\bibnamefont {Diot}},\ and\ \bibinfo
  {author} {\bibfnamefont {M.}~\bibnamefont {Prentiss}},\ }\bibfield  {title}
  {\bibinfo {title} {Splitting matter waves using an optimized standing-wave
  light-pulse sequence},\ }\href {https://doi.org/10.1103/PhysRevA.71.043602}
  {\bibfield  {journal} {\bibinfo  {journal} {Physical Review A}\ }\textbf
  {\bibinfo {volume} {71}},\ \bibinfo {pages} {043602} (\bibinfo {year}
  {2005})}\BibitemShut {NoStop}%
\bibitem [{\citenamefont {Niu}\ \emph {et~al.}(2018)\citenamefont {Niu},
  \citenamefont {Guo}, \citenamefont {Zhan}, \citenamefont {Chen},
  \citenamefont {Liu},\ and\ \citenamefont {Zhou}}]{niu2018optimized}%
  \BibitemOpen
  \bibfield  {author} {\bibinfo {author} {\bibfnamefont {L.}~\bibnamefont
  {Niu}}, \bibinfo {author} {\bibfnamefont {X.}~\bibnamefont {Guo}}, \bibinfo
  {author} {\bibfnamefont {Y.}~\bibnamefont {Zhan}}, \bibinfo {author}
  {\bibfnamefont {X.}~\bibnamefont {Chen}}, \bibinfo {author} {\bibfnamefont
  {W.~M.}\ \bibnamefont {Liu}},\ and\ \bibinfo {author} {\bibfnamefont
  {X.}~\bibnamefont {Zhou}},\ }\bibfield  {title} {\bibinfo {title} {Optimized
  fringe removal algorithm for absorption images},\ }\href
  {https://doi.org/10.1063/1.5040669} {\bibfield  {journal} {\bibinfo
  {journal} {Applied Physics Letters}\ }\textbf {\bibinfo {volume} {113}},\
  \bibinfo {pages} {144103} (\bibinfo {year} {2018})}\BibitemShut {NoStop}%
\bibitem [{\citenamefont {Ockeloen}\ \emph {et~al.}(2010)\citenamefont
  {Ockeloen}, \citenamefont {Tauschinsky}, \citenamefont {Spreeuw},\ and\
  \citenamefont {Whitlock}}]{ockeloen2010detection}%
  \BibitemOpen
  \bibfield  {author} {\bibinfo {author} {\bibfnamefont {C.~F.}\ \bibnamefont
  {Ockeloen}}, \bibinfo {author} {\bibfnamefont {A.~F.}\ \bibnamefont
  {Tauschinsky}}, \bibinfo {author} {\bibfnamefont {R.~J.~C.}\ \bibnamefont
  {Spreeuw}},\ and\ \bibinfo {author} {\bibfnamefont {S.}~\bibnamefont
  {Whitlock}},\ }\bibfield  {title} {\bibinfo {title} {Detection of small atom
  numbers through image processing},\ }\href
  {https://doi.org/10.1103/PhysRevA.82.061606} {\bibfield  {journal} {\bibinfo
  {journal} {Physical Review A}\ }\textbf {\bibinfo {volume} {82}},\ \bibinfo
  {pages} {061606} (\bibinfo {year} {2010})}\BibitemShut {NoStop}%
\bibitem [{\citenamefont {Peters}\ \emph {et~al.}(2001)\citenamefont {Peters},
  \citenamefont {Chung},\ and\ \citenamefont {Chu}}]{peters2001high}%
  \BibitemOpen
  \bibfield  {author} {\bibinfo {author} {\bibfnamefont {A.}~\bibnamefont
  {Peters}}, \bibinfo {author} {\bibfnamefont {K.~Y.}\ \bibnamefont {Chung}},\
  and\ \bibinfo {author} {\bibfnamefont {S.}~\bibnamefont {Chu}},\ }\bibfield
  {title} {\bibinfo {title} {High-precision gravity measurements using atom
  interferometry},\ }\href {https://doi.org/10.1088/0026-1394/38/1/4}
  {\bibfield  {journal} {\bibinfo  {journal} {Metrologia}\ }\textbf {\bibinfo
  {volume} {38}},\ \bibinfo {pages} {25} (\bibinfo {year} {2001})}\BibitemShut
  {NoStop}%
\bibitem [{\citenamefont {Gustavsson}\ \emph {et~al.}(2008)\citenamefont
  {Gustavsson}, \citenamefont {Haller}, \citenamefont {Mark}, \citenamefont
  {Danzl}, \citenamefont {Rojas-Kopeinig},\ and\ \citenamefont
  {Nägerl}}]{gustavsson2008control}%
  \BibitemOpen
  \bibfield  {author} {\bibinfo {author} {\bibfnamefont {M.}~\bibnamefont
  {Gustavsson}}, \bibinfo {author} {\bibfnamefont {E.}~\bibnamefont {Haller}},
  \bibinfo {author} {\bibfnamefont {M.~J.}\ \bibnamefont {Mark}}, \bibinfo
  {author} {\bibfnamefont {J.~G.}\ \bibnamefont {Danzl}}, \bibinfo {author}
  {\bibfnamefont {G.}~\bibnamefont {Rojas-Kopeinig}},\ and\ \bibinfo {author}
  {\bibfnamefont {H.~C.}\ \bibnamefont {Nägerl}},\ }\bibfield  {title}
  {\bibinfo {title} {Control of interaction-induced dephasing of bloch
  oscillations},\ }\href {https://doi.org/10.1103/PhysRevLett.100.080404}
  {\bibfield  {journal} {\bibinfo  {journal} {Physical Review Letters}\
  }\textbf {\bibinfo {volume} {100}},\ \bibinfo {pages} {080404} (\bibinfo
  {year} {2008})}\BibitemShut {NoStop}%
\bibitem [{\citenamefont {McDonald}\ \emph {et~al.}(2014)\citenamefont
  {McDonald}, \citenamefont {Kuhn}, \citenamefont {Hardman}, \citenamefont
  {Bennetts}, \citenamefont {Everitt}, \citenamefont {Altin}, \citenamefont
  {Debs}, \citenamefont {Close},\ and\ \citenamefont
  {Robins}}]{mcdonald2014bright}%
  \BibitemOpen
  \bibfield  {author} {\bibinfo {author} {\bibfnamefont {G.~D.}\ \bibnamefont
  {McDonald}}, \bibinfo {author} {\bibfnamefont {C.~C.~N.}\ \bibnamefont
  {Kuhn}}, \bibinfo {author} {\bibfnamefont {K.~S.}\ \bibnamefont {Hardman}},
  \bibinfo {author} {\bibfnamefont {S.}~\bibnamefont {Bennetts}}, \bibinfo
  {author} {\bibfnamefont {P.~J.}\ \bibnamefont {Everitt}}, \bibinfo {author}
  {\bibfnamefont {P.~A.}\ \bibnamefont {Altin}}, \bibinfo {author}
  {\bibfnamefont {J.~E.}\ \bibnamefont {Debs}}, \bibinfo {author}
  {\bibfnamefont {J.~D.}\ \bibnamefont {Close}},\ and\ \bibinfo {author}
  {\bibfnamefont {N.~P.}\ \bibnamefont {Robins}},\ }\bibfield  {title}
  {\bibinfo {title} {Bright solitonic matter-wave interferometer},\ }\href
  {http://link.aps.org/doi/10.1103/PhysRevLett.113.013002} {\bibfield
  {journal} {\bibinfo  {journal} {Physical Review Letters}\ }\textbf {\bibinfo
  {volume} {113}},\ \bibinfo {pages} {013002} (\bibinfo {year}
  {2014})}\BibitemShut {NoStop}%
\bibitem [{\citenamefont {Yuge}\ \emph {et~al.}(2011)\citenamefont {Yuge},
  \citenamefont {Sasaki},\ and\ \citenamefont
  {Hirayama}}]{yuge2011measurement}%
  \BibitemOpen
  \bibfield  {author} {\bibinfo {author} {\bibfnamefont {T.}~\bibnamefont
  {Yuge}}, \bibinfo {author} {\bibfnamefont {S.}~\bibnamefont {Sasaki}},\ and\
  \bibinfo {author} {\bibfnamefont {Y.}~\bibnamefont {Hirayama}},\ }\bibfield
  {title} {\bibinfo {title} {Measurement of the noise spectrum using a
  multiple-pulse sequence},\ }\href
  {https://doi.org/10.1103/PhysRevLett.107.170504} {\bibfield  {journal}
  {\bibinfo  {journal} {Physical Review Letters}\ }\textbf {\bibinfo {volume}
  {107}},\ \bibinfo {pages} {170504} (\bibinfo {year} {2011})}\BibitemShut
  {NoStop}%
\bibitem [{\citenamefont {Deissler}\ \emph {et~al.}(2008)\citenamefont
  {Deissler}, \citenamefont {Hughes}, \citenamefont {Burke},\ and\
  \citenamefont {Sackett}}]{deissler2008measurement}%
  \BibitemOpen
  \bibfield  {author} {\bibinfo {author} {\bibfnamefont {B.}~\bibnamefont
  {Deissler}}, \bibinfo {author} {\bibfnamefont {K.~J.}\ \bibnamefont
  {Hughes}}, \bibinfo {author} {\bibfnamefont {J.~H.~T.}\ \bibnamefont
  {Burke}},\ and\ \bibinfo {author} {\bibfnamefont {C.~A.}\ \bibnamefont
  {Sackett}},\ }\bibfield  {title} {\bibinfo {title} {Measurement of the ac
  {Stark} shift with a guided matter-wave interferometer},\ }\href
  {http://link.aps.org/doi/10.1103/PhysRevA.77.031604} {\bibfield  {journal}
  {\bibinfo  {journal} {Physical Review A}\ }\textbf {\bibinfo {volume} {77}},\
  \bibinfo {pages} {031604} (\bibinfo {year} {2008})}\BibitemShut {NoStop}%
\bibitem [{\citenamefont {Parinaz}\ \emph {et~al.}(2021)\citenamefont
  {Parinaz}, \citenamefont {Adam}, \citenamefont {Bindiya}, \citenamefont
  {Rudolf},\ and\ \citenamefont {Marianna}}]{parinaz2021portal}%
  \BibitemOpen
  \bibfield  {author} {\bibinfo {author} {\bibfnamefont {B.}~\bibnamefont
  {Parinaz}}, \bibinfo {author} {\bibfnamefont {M.}~\bibnamefont {Adam}},
  \bibinfo {author} {\bibfnamefont {A.}~\bibnamefont {Bindiya}}, \bibinfo
  {author} {\bibfnamefont {E.}~\bibnamefont {Rudolf}},\ and\ \bibinfo {author}
  {\bibfnamefont {S.~S.}\ \bibnamefont {Marianna}},\ }\href
  {https://www.udel.edu/atom} {\bibinfo {title} {Portal for high-precision
  atomic data and computation (version 1.0)}} (\bibinfo {year}
  {2021})\BibitemShut {NoStop}%
\bibitem [{\citenamefont {Krzyzanowska}\ \emph {et~al.}(2022)\citenamefont
  {Krzyzanowska}, \citenamefont {Ferreras}, \citenamefont {Ryu}, \citenamefont
  {Samson},\ and\ \citenamefont {Boshier}}]{krzyzanowska2022matter}%
  \BibitemOpen
  \bibfield  {author} {\bibinfo {author} {\bibfnamefont {K.}~\bibnamefont
  {Krzyzanowska}}, \bibinfo {author} {\bibfnamefont {J.}~\bibnamefont
  {Ferreras}}, \bibinfo {author} {\bibfnamefont {C.}~\bibnamefont {Ryu}},
  \bibinfo {author} {\bibfnamefont {C.}~\bibnamefont {Samson}},\ and\ \bibinfo
  {author} {\bibfnamefont {M.}~\bibnamefont {Boshier}},\ }\bibfield  {title}
  {\bibinfo {title} {Matter wave analog of a fiber optic gyro},\ }\Eprint
  {https://arxiv.org/abs/2201.12461} {arXiv:2201.12461 [physics.atom-ph]}
  (\bibinfo {year} {2022})\BibitemShut {NoStop}%
\bibitem [{\citenamefont {Pandey}\ \emph {et~al.}(2019)\citenamefont {Pandey},
  \citenamefont {Mas}, \citenamefont {Drougakis}, \citenamefont {Thekkeppatt},
  \citenamefont {Bolpasi}, \citenamefont {Vasilakis}, \citenamefont {Poulios},\
  and\ \citenamefont {von Klitzing}}]{pandey2019hypersonic}%
  \BibitemOpen
  \bibfield  {author} {\bibinfo {author} {\bibfnamefont {S.}~\bibnamefont
  {Pandey}}, \bibinfo {author} {\bibfnamefont {H.}~\bibnamefont {Mas}},
  \bibinfo {author} {\bibfnamefont {G.}~\bibnamefont {Drougakis}}, \bibinfo
  {author} {\bibfnamefont {P.}~\bibnamefont {Thekkeppatt}}, \bibinfo {author}
  {\bibfnamefont {V.}~\bibnamefont {Bolpasi}}, \bibinfo {author} {\bibfnamefont
  {G.}~\bibnamefont {Vasilakis}}, \bibinfo {author} {\bibfnamefont
  {K.}~\bibnamefont {Poulios}},\ and\ \bibinfo {author} {\bibfnamefont
  {W.}~\bibnamefont {von Klitzing}},\ }\bibfield  {title} {\bibinfo {title}
  {Hypersonic {Bose–Einstein} condensates in accelerator rings},\ }\href
  {https://doi.org/10.1038/s41586-019-1273-5} {\bibfield  {journal} {\bibinfo
  {journal} {Nature}\ }\textbf {\bibinfo {volume} {570}},\ \bibinfo {pages}
  {205} (\bibinfo {year} {2019})}\BibitemShut {NoStop}%
\bibitem [{\citenamefont {Moan}\ \emph {et~al.}(2020)\citenamefont {Moan},
  \citenamefont {Horne}, \citenamefont {Arpornthip}, \citenamefont {Luo},
  \citenamefont {Fallon}, \citenamefont {Berl},\ and\ \citenamefont
  {Sackett}}]{moan2020quantum}%
  \BibitemOpen
  \bibfield  {author} {\bibinfo {author} {\bibfnamefont {E.~R.}\ \bibnamefont
  {Moan}}, \bibinfo {author} {\bibfnamefont {R.~A.}\ \bibnamefont {Horne}},
  \bibinfo {author} {\bibfnamefont {T.}~\bibnamefont {Arpornthip}}, \bibinfo
  {author} {\bibfnamefont {Z.}~\bibnamefont {Luo}}, \bibinfo {author}
  {\bibfnamefont {A.~J.}\ \bibnamefont {Fallon}}, \bibinfo {author}
  {\bibfnamefont {S.~J.}\ \bibnamefont {Berl}},\ and\ \bibinfo {author}
  {\bibfnamefont {C.~A.}\ \bibnamefont {Sackett}},\ }\bibfield  {title}
  {\bibinfo {title} {Quantum rotation sensing with dual {Sagnac}
  interferometers in an atom-optical waveguide},\ }\href
  {https://doi.org/10.1103/PhysRevLett.124.120403} {\bibfield  {journal}
  {\bibinfo  {journal} {Physical Review Letters}\ }\textbf {\bibinfo {volume}
  {124}},\ \bibinfo {pages} {120403} (\bibinfo {year} {2020})}\BibitemShut
  {NoStop}%
\bibitem [{\citenamefont {Savoie}\ \emph {et~al.}(2018)\citenamefont {Savoie},
  \citenamefont {Altorio}, \citenamefont {Fang}, \citenamefont {Sidorenkov},
  \citenamefont {Geiger},\ and\ \citenamefont
  {Landragin}}]{savoie2018iinterleaved}%
  \BibitemOpen
  \bibfield  {author} {\bibinfo {author} {\bibfnamefont {D.}~\bibnamefont
  {Savoie}}, \bibinfo {author} {\bibfnamefont {M.}~\bibnamefont {Altorio}},
  \bibinfo {author} {\bibfnamefont {B.}~\bibnamefont {Fang}}, \bibinfo {author}
  {\bibfnamefont {L.~A.}\ \bibnamefont {Sidorenkov}}, \bibinfo {author}
  {\bibfnamefont {R.}~\bibnamefont {Geiger}},\ and\ \bibinfo {author}
  {\bibfnamefont {A.}~\bibnamefont {Landragin}},\ }\bibfield  {title} {\bibinfo
  {title} {Interleaved atom interferometry for high-sensitivity inertial
  measurements},\ }\href {https://doi.org/10.1126/sciadv.aau7948} {\bibfield
  {journal} {\bibinfo  {journal} {Science Advances}\ }\textbf {\bibinfo
  {volume} {4}},\ \bibinfo {pages} {eaau7948} (\bibinfo {year}
  {2018})}\BibitemShut {NoStop}%
\bibitem [{\citenamefont {Graham}\ \emph {et~al.}(2016)\citenamefont {Graham},
  \citenamefont {Hogan}, \citenamefont {Kasevich},\ and\ \citenamefont
  {Rajendran}}]{graham2016resonant}%
  \BibitemOpen
  \bibfield  {author} {\bibinfo {author} {\bibfnamefont {P.~W.}\ \bibnamefont
  {Graham}}, \bibinfo {author} {\bibfnamefont {J.~M.}\ \bibnamefont {Hogan}},
  \bibinfo {author} {\bibfnamefont {M.~A.}\ \bibnamefont {Kasevich}},\ and\
  \bibinfo {author} {\bibfnamefont {S.}~\bibnamefont {Rajendran}},\ }\bibfield
  {title} {\bibinfo {title} {Resonant mode for gravitational wave detectors
  based on atom interferometry},\ }\href
  {https://doi.org/10.1103/PhysRevD.94.104022} {\bibfield  {journal} {\bibinfo
  {journal} {Physical Review D}\ }\textbf {\bibinfo {volume} {94}},\ \bibinfo
  {pages} {104022} (\bibinfo {year} {2016})}\BibitemShut {NoStop}%
\bibitem [{\citenamefont {Landini}\ \emph {et~al.}(2012)\citenamefont
  {Landini}, \citenamefont {Roy}, \citenamefont {Roati}, \citenamefont
  {Simoni}, \citenamefont {Inguscio}, \citenamefont {Modugno},\ and\
  \citenamefont {Fattori}}]{landini2012direct}%
  \BibitemOpen
  \bibfield  {author} {\bibinfo {author} {\bibfnamefont {M.}~\bibnamefont
  {Landini}}, \bibinfo {author} {\bibfnamefont {S.}~\bibnamefont {Roy}},
  \bibinfo {author} {\bibfnamefont {G.}~\bibnamefont {Roati}}, \bibinfo
  {author} {\bibfnamefont {A.}~\bibnamefont {Simoni}}, \bibinfo {author}
  {\bibfnamefont {M.}~\bibnamefont {Inguscio}}, \bibinfo {author}
  {\bibfnamefont {G.}~\bibnamefont {Modugno}},\ and\ \bibinfo {author}
  {\bibfnamefont {M.}~\bibnamefont {Fattori}},\ }\bibfield  {title} {\bibinfo
  {title} {Direct evaporative cooling of $^{39}\text{K}$ atoms to
  {Bose-Einstein} condensation},\ }\href
  {https://doi.org/10.1103/PhysRevA.86.033421} {\bibfield  {journal} {\bibinfo
  {journal} {Physical Review A}\ }\textbf {\bibinfo {volume} {86}},\ \bibinfo
  {pages} {033421} (\bibinfo {year} {2012})}\BibitemShut {NoStop}%
\bibitem [{\citenamefont {Salomon}\ \emph {et~al.}(2013)\citenamefont
  {Salomon}, \citenamefont {Fouché}, \citenamefont {Wang}, \citenamefont
  {Aspect}, \citenamefont {Bouyer},\ and\ \citenamefont
  {Bourdel}}]{salomon2013gray}%
  \BibitemOpen
  \bibfield  {author} {\bibinfo {author} {\bibfnamefont {G.}~\bibnamefont
  {Salomon}}, \bibinfo {author} {\bibfnamefont {L.}~\bibnamefont {Fouché}},
  \bibinfo {author} {\bibfnamefont {P.}~\bibnamefont {Wang}}, \bibinfo {author}
  {\bibfnamefont {A.}~\bibnamefont {Aspect}}, \bibinfo {author} {\bibfnamefont
  {P.}~\bibnamefont {Bouyer}},\ and\ \bibinfo {author} {\bibfnamefont
  {T.}~\bibnamefont {Bourdel}},\ }\bibfield  {title} {\bibinfo {title}
  {Gray-molasses cooling of $^{39}\text{K}$ to a high phase-space density},\
  }\href {https://doi.org/10.1209/0295-5075/104/63002} {\bibfield  {journal}
  {\bibinfo  {journal} {EPL (Europhysics Letters)}\ }\textbf {\bibinfo {volume}
  {104}},\ \bibinfo {pages} {63002} (\bibinfo {year} {2013})}\BibitemShut
  {NoStop}%
\bibitem [{\citenamefont {Couvert}\ \emph {et~al.}(2008)\citenamefont
  {Couvert}, \citenamefont {Kawalec}, \citenamefont {Reinaudi},\ and\
  \citenamefont {Guéry-Odelin}}]{couvert2008optimal}%
  \BibitemOpen
  \bibfield  {author} {\bibinfo {author} {\bibfnamefont {A.}~\bibnamefont
  {Couvert}}, \bibinfo {author} {\bibfnamefont {T.}~\bibnamefont {Kawalec}},
  \bibinfo {author} {\bibfnamefont {G.}~\bibnamefont {Reinaudi}},\ and\
  \bibinfo {author} {\bibfnamefont {D.}~\bibnamefont {Guéry-Odelin}},\
  }\bibfield  {title} {\bibinfo {title} {Optimal transport of ultracold atoms
  in the non-adiabatic regime},\ }\href
  {https://doi.org/10.1209/0295-5075/83/13001} {\bibfield  {journal} {\bibinfo
  {journal} {EPL (Europhysics Letters)}\ }\textbf {\bibinfo {volume} {83}},\
  \bibinfo {pages} {13001} (\bibinfo {year} {2008})}\BibitemShut {NoStop}%
\bibitem [{\citenamefont {Gross}\ \emph {et~al.}(2016)\citenamefont {Gross},
  \citenamefont {Gan},\ and\ \citenamefont {Dieckmann}}]{gross2016all}%
  \BibitemOpen
  \bibfield  {author} {\bibinfo {author} {\bibfnamefont {C.}~\bibnamefont
  {Gross}}, \bibinfo {author} {\bibfnamefont {H.~C.~J.}\ \bibnamefont {Gan}},\
  and\ \bibinfo {author} {\bibfnamefont {K.}~\bibnamefont {Dieckmann}},\
  }\bibfield  {title} {\bibinfo {title} {All-optical production and transport
  of a large $^{6}\mathrm{Li}$ quantum gas in a crossed optical dipole trap},\
  }\href {https://doi.org/10.1103/PhysRevA.93.053424} {\bibfield  {journal}
  {\bibinfo  {journal} {Physical Review A}\ }\textbf {\bibinfo {volume} {93}},\
  \bibinfo {pages} {053424} (\bibinfo {year} {2016})}\BibitemShut {NoStop}%
\bibitem [{\citenamefont {Stamper-Kurn}\ \emph {et~al.}(1998)\citenamefont
  {Stamper-Kurn}, \citenamefont {Miesner}, \citenamefont {Chikkatur},
  \citenamefont {Inouye}, \citenamefont {Stenger},\ and\ \citenamefont
  {Ketterle}}]{stamperkurn1998reversible}%
  \BibitemOpen
  \bibfield  {author} {\bibinfo {author} {\bibfnamefont {D.~M.}\ \bibnamefont
  {Stamper-Kurn}}, \bibinfo {author} {\bibfnamefont {H.~J.}\ \bibnamefont
  {Miesner}}, \bibinfo {author} {\bibfnamefont {A.~P.}\ \bibnamefont
  {Chikkatur}}, \bibinfo {author} {\bibfnamefont {S.}~\bibnamefont {Inouye}},
  \bibinfo {author} {\bibfnamefont {J.}~\bibnamefont {Stenger}},\ and\ \bibinfo
  {author} {\bibfnamefont {W.}~\bibnamefont {Ketterle}},\ }\bibfield  {title}
  {\bibinfo {title} {Reversible formation of a {Bose-Einstein} condensate},\
  }\href {https://doi.org/10.1103/PhysRevLett.81.2194} {\bibfield  {journal}
  {\bibinfo  {journal} {Physical Review Letters}\ }\textbf {\bibinfo {volume}
  {81}},\ \bibinfo {pages} {2194} (\bibinfo {year} {1998})}\BibitemShut
  {NoStop}%
\bibitem [{\citenamefont {Pinkse}\ \emph {et~al.}(1997)\citenamefont {Pinkse},
  \citenamefont {Mosk}, \citenamefont {Weidem\"{u}ller}, \citenamefont
  {Reynolds}, \citenamefont {Hijmans},\ and\ \citenamefont
  {Walraven}}]{pinkse1997adiabtically}%
  \BibitemOpen
  \bibfield  {author} {\bibinfo {author} {\bibfnamefont {P.~W.~H.}\
  \bibnamefont {Pinkse}}, \bibinfo {author} {\bibfnamefont {A.}~\bibnamefont
  {Mosk}}, \bibinfo {author} {\bibfnamefont {M.}~\bibnamefont
  {Weidem\"{u}ller}}, \bibinfo {author} {\bibfnamefont {M.~W.}\ \bibnamefont
  {Reynolds}}, \bibinfo {author} {\bibfnamefont {T.~W.}\ \bibnamefont
  {Hijmans}},\ and\ \bibinfo {author} {\bibfnamefont {J.~T.~M.}\ \bibnamefont
  {Walraven}},\ }\bibfield  {title} {\bibinfo {title} {Adiabatically changing
  the phase-space density of a trapped {Bose} gas},\ }\href
  {https://doi.org/10.1103/PhysRevLett.78.990} {\bibfield  {journal} {\bibinfo
  {journal} {Physical Review Letters}\ }\textbf {\bibinfo {volume} {78}},\
  \bibinfo {pages} {990} (\bibinfo {year} {1997})}\BibitemShut {NoStop}%
\bibitem [{\citenamefont {O’Hara}\ \emph {et~al.}(2001)\citenamefont
  {O’Hara}, \citenamefont {Gehm}, \citenamefont {Granade},\ and\
  \citenamefont {Thomas}}]{ohara2001scaling}%
  \BibitemOpen
  \bibfield  {author} {\bibinfo {author} {\bibfnamefont {K.~M.}\ \bibnamefont
  {O’Hara}}, \bibinfo {author} {\bibfnamefont {M.~E.}\ \bibnamefont {Gehm}},
  \bibinfo {author} {\bibfnamefont {S.~R.}\ \bibnamefont {Granade}},\ and\
  \bibinfo {author} {\bibfnamefont {J.~E.}\ \bibnamefont {Thomas}},\ }\bibfield
   {title} {\bibinfo {title} {Scaling laws for evaporative cooling in
  time-dependent optical traps},\ }\href
  {https://doi.org/10.1103/PhysRevA.64.051403} {\bibfield  {journal} {\bibinfo
  {journal} {Physical Review A}\ }\textbf {\bibinfo {volume} {64}},\ \bibinfo
  {pages} {051403} (\bibinfo {year} {2001})}\BibitemShut {NoStop}%
\bibitem [{\citenamefont {Crosser}\ \emph {et~al.}(2010)\citenamefont
  {Crosser}, \citenamefont {Scott}, \citenamefont {Clark},\ and\ \citenamefont
  {Wilt}}]{crosser2010on}%
  \BibitemOpen
  \bibfield  {author} {\bibinfo {author} {\bibfnamefont {M.~S.}\ \bibnamefont
  {Crosser}}, \bibinfo {author} {\bibfnamefont {S.}~\bibnamefont {Scott}},
  \bibinfo {author} {\bibfnamefont {A.}~\bibnamefont {Clark}},\ and\ \bibinfo
  {author} {\bibfnamefont {P.~M.}\ \bibnamefont {Wilt}},\ }\bibfield  {title}
  {\bibinfo {title} {On the magnetic field near the center of {Helmholtz}
  coils},\ }\href {https://doi.org/10.1063/1.3474227} {\bibfield  {journal}
  {\bibinfo  {journal} {Review of Scientific Instruments}\ }\textbf {\bibinfo
  {volume} {81}},\ \bibinfo {pages} {084701} (\bibinfo {year}
  {2010})}\BibitemShut {NoStop}%
\end{thebibliography}%

\pagebreak
\section*{Supplemental Material}
\definecolor{apsBlue}{RGB}{50, 50, 150}


\newcommand{\figExperimentalScheme}{1}
\newcommand{\figInterrogationTimeLimits}{2}
\newcommand{\figFringes}{3}

\title{Supplemental Material}
\preprint{LA-UR-22-20692}

\maketitle

\setcounter{equation}{0}
\renewcommand{\theequation}{S\arabic{equation}}
\setcounter{figure}{0}
\renewcommand{\thefigure}{S\arabic{figure}}

\section{BEC preparation}

\begin{figure}[b]
\includegraphics[width=1\columnwidth]{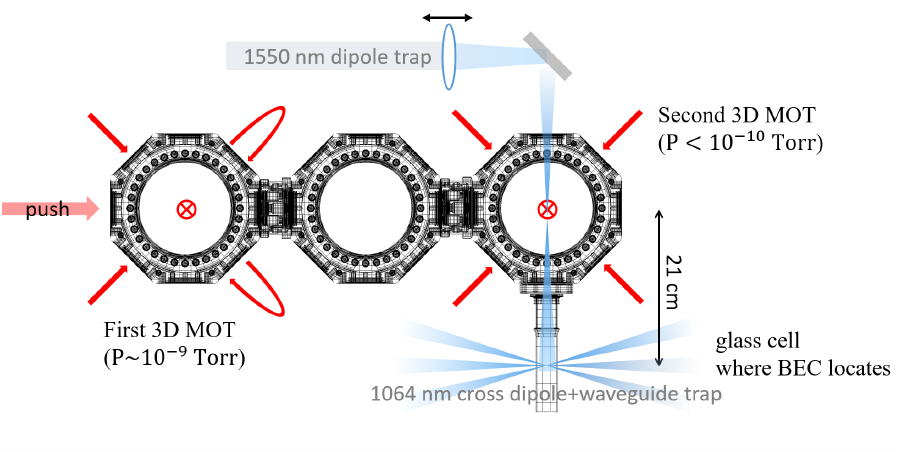} 
\caption{\label{fig:ThreeChambers} Experimental setup of three-stage BEC machine.}
\end{figure}

Preparation of nearly pure BEC for the experiment discussed in this paper is achieved in three stages. As shown in Fig.~\ref{fig:ThreeChambers}, the apparatus consists of two vacuum chambers plus a glass cell where the BEC is prepared. The first stage takes place in a relatively high pressure chamber where $^{39}$K cold atoms are continuously loaded into a 3D $D$2-line magneto-optical trap (MOT) from a vapor provided by commercial getter sources.
This high pressure chamber is separated from the rest of the system by a constriction and an empty chamber with an ion pump, enabling differential pumping. Simultaneously, a near-resonance push beam ($D$2 cooling plus repumping light) overlapped with the high pressure MOT continuously pushes atoms towards the second MOT in the low pressure chamber. Atoms are collected in the low pressure MOT for about 5~s, with a loading rate $\sim10^8~$atoms/s.

Once the atomic cloud is formed in the low pressure chamber, the push beam is blocked and the second stage of the preparation begins. It starts with $50~$ms $D$2 compression with field gradient $24~$G/cm \cite{landini2012direct}, followed by $10~$ms of $D$1-$D$2 hybrid compression. At this stage the peak density of the atomic cloud is $\sim8\times10^{10}~$atoms/cm$^3$ with $60~\mu$K temperature. Next, the temperature is reduced to $6~\mu$K using $D$1 gray molasses technique with the magnetic field gradient set to near zero for $7~$ms~\cite{salomon2013gray}. The atoms are optically pumped into the $F=1$ manifold by turning off the $D$1 repump light for the last $3~$ms of gray molasses. The resulting atoms are then captured in a magnetic trap by turning on the MOT anti-Helmholtz coils at a gradient of $44~$G/cm which is then ramped linearly to $200~$G/cm over $100~$ms. Confinement in a tight magnetic trap provides the time needed to overlap a dipole trap beam with the atomic cloud for about 2~s and load a sufficiently high fraction of atoms into the beam. 

The dipole trap beam is a 1550~nm single mode laser (IPG photonics, ELR-LP-SF-30) with a total power 20~W and active power stabilization. The beam is focused by a singlet lens with $f=750~$mm and delivered to the low pressure chamber, providing $720~\mu$K  trap depth and $2\pi\times36~$Hz axial trap frequency. Importantly, the dipole trap is slightly displaced from the center of the magnetic trap to hold the spin polarized atoms. Once the dipole trap is loaded from the magnetic trap, the magnetic field is turned off. A typical atom number density and temperature in the dipole trap obtained at this point are $\sim10^6,10^{11}\text{cm}^{-3}$ and $15~\mu$K respectively.

In the final stage of the preparation, the atoms are transferred to the science cell by translating the lens on a linear motor stage (Newport, XML350) at a distance $215~$mm during a $900~$ms trajectory~\cite{couvert2008optimal, gross2016all}. The velocity profile of the translation stage has isosceles trapezoid shape of maximum velocity $280~$mm/s and acceleration $2200~$mm/s$^2$. The transfer efficiency is close to $100\%$ without any discernable heating. The transfer time had to be increased to a few seconds recently because wear in the stage after a million transport cycles increased vibration during transport. The Earth's magnetic field maintains the spin polarization during the transfer. Light-assisted inelastic loss is negligible during the transfer because the trap depth is near the Ramsauer minimum~\cite{landini2012direct}. 

The transition from thermal cloud to a BEC takes place in the science cell. An additional set of 1064~nm beams is overlapped with the atomic cloud held in 1550~nm dipole trap, as shown in Fig.~\ref{fig:ThreeChambers}. The s-wave scattering length is set to $200a_0$ by the magnetic Feshbach resonance at $562~$G~\cite{derrico2007feshbach}, and the power of the 1550~nm dipole trap is reduced for $1.3$s to load the atoms into the 1064~nm cross-dipole trap by the dimple effect~\cite{stamperkurn1998reversible, pinkse1997adiabtically}. For the next $200~$ms an adiabatic compression is applied in the 1064~nm beams by increasing the power by factor of 2 and setting the s-wave scattering length to near zero. Finally, a 5~s step of forced evaporative cooling takes place, when the scattering length is set to $\sim500a_0$ and the beams' intensity is ramped~\cite{ohara2001scaling} to a final trap frequency $\omega_{x,y,z}=2\pi(77,100,12)~$Hz.

\begin{figure*}
\includegraphics[width=1\textwidth]{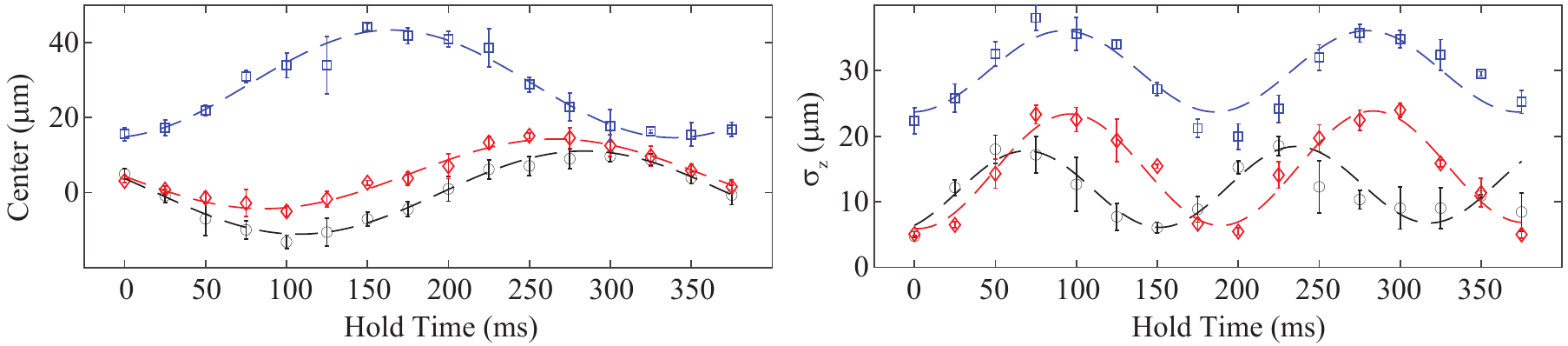} 
\caption{\label{fig:BecMotion} Center position (left) and axial width $\sigma_z$ (right) of BEC after $16~$ms free expansion, as a function of time in the waveguide. Dashed lines are sinusoidal fits. Error bars are standard error in the mean of at least 4 measurements. (Black circles, red diamonds, blue squares) correspond respectively to the $s$-wave scattering length values $a_s=(-2.5, 0.7, 150)a_0$}
\end{figure*}

Next, as shown in Fig.~\ref{fig:ThreeChambers}, the outer two 1064~nm beams in the crossed dipole trap are slowly turned off, transferring the BEC with minimal excitation to the middle ($z$-axis in Fig.~\figExperimentalScheme) 1064~nm trapping beam which forms the waveguide. The z-axis harmonic potential ($\omega_z=2\pi\times2.8~Hz$) is almost entirely due to the inhomogeneity of the field produced by the Helmholtz coils~\cite{crosser2010on}; the $2\pi\times$0.3~Hz axial trapping frequency of the guide is negligible. The mismatch in waveguide and trapping potentials excites both center-of-mass motion and a compressional mode (discussed below), but importantly the wavepacket does not disperse during the interferometer interrogation time. Finally, the s-wave scattering length is then set to a desired value by linearly sweeping the magnetic field for $300~$ms. The typical BEC preparation time to this point is about 20 s.

\section{Interferometer sequence}
The momentum state of the BEC is manipulated using Bragg scattering from a standing wave of light formed by a 766.5~nm external cavity diode laser detuned 100~GHz blue of the $D$2 line and delivered to the interaction region with a polarization-maintaining optical fiber and switched with an acousto-optic modulator. The standing wave is created by retro-reflection from a dichroic mirror (Fig.~\figExperimentalScheme(a)) of a beam with waist ($1/e^2$) at the atoms of $80~\mu$m to prevent interference from overlap of internal reflections from the surfaces of the uncoated glass cell. Sub-ms pulses of the standing wave grating impart $\pm2\hbar k$ momentum kicks, where $k$ is the wave-vector of the Bragg laser~\cite{szigeti2012why,muller2008atom,wu2005splitting}. As shown in Fig.~\figExperimentalScheme(b), a double square pulse~\cite{wu2005splitting} acts as a beamsplitter that transforms the initially stationary atoms into a coherent superposition of two momentum states, $(\psi_++\psi_-)/\sqrt{2}$. Subsequent gaussian pulses~\cite{muller2008atom} act as mirrors, reversing the momentum of the wavepackets, $\psi_+\leftrightarrow\psi_-$. At the end of the cycle a second beamsplitter pulse with adjustable laser phase $\theta$ relative to the previous pulses recombines the atoms into a superposition of the two output ports of the interferometer. The waveguide and the magnetic field are then turned off, and the atom numbers in each of the momentum states ($N_{+,-,0}$) are measured by absorption imaging after $16~$ms of free expansion. Background noise in the absorption images is reduced with a fringe suppression algorithm~\cite{ockeloen2010detection, niu2018optimized}. We note that the low spontaneous scattering rate of photons from the Bragg laser ($2\pi\times$ 0.03~Hz) means that process has negligible effect on the interferometer. 

The relative phase of the recombination pulse $\theta$ is tuned by changing the frequency of Bragg laser ($\delta f$) using its built-in piezo-electric tuning element. The resulting phase shift is $\theta=8\pi\times\delta f\times d/c$, where $d$ is the distance from the mirror to the atoms and $c$ is the speed of light. In our system where $d\sim$~0.4~m, a frequency shift $\delta f\sim$100~MHz is enough to change $\theta$ by $\pi$. The intensity and frequency of the Bragg laser are not actively stabilized. Nor are the position and tilt of the retro-reflector.

\section{BEC dynamics during the interferometer sequence}
The transfer of the BEC from the crossed-dipole trap to the waveguide beam is designed to be adiabatic. However, imperfections in the system induce a center-of-mass oscillation and excite a compressional mode of the BEC. The resulting dynamics are shown in Fig.~\ref{fig:BecMotion}, where the center-of-mass position (left) and the axial width (right) of BEC are plotted as a function of the hold time in the waveguide. The compressional mode frequency $\omega_c$ and the axial trap frequency  $\omega_z=2\pi\times[2.8(1), 2.8(1), 2.9(2)]~$Hz are found from sinusoidal fits for scattering lengths $a_s=[-2.5, 0.7, 150]a_0$ respectively.The axial trap frequency of the guide ($2\pi\times$0.3~Hz) is negligible; it is the z-axis harmonic potential ($\omega_z=2\pi\times2.8~$Hz) from the inhomogeneity of the Helmholtz coils~\cite{crosser2010on} which induces the compressional mode ~\cite{dicarli2019excitation, haller2009realization}.

The initial center-of-mass momentum can be estimated as $p_{initial}=m\omega_zz_{initial}$, where $z_{initial}$ is the amplitude of the sinusoidal fit. The initial momentum depends on the details of the final steps in the preparation sequence, namely the end of the evaporation and the loading into the waveguide. The center-of-mass position changes as a function of the final s-wave scattering length because the potential minimum moves when the magnetic trap depth is reduced. Since the the BEC is located near the waveguide beam waist, by comparing the minimum position of the combined potential $m(\omega_zz)^2/2+mgz\theta_{tilt}$ to the measured center-of-mass position for various $\omega_z$, we can  estimate that the tilt angle between the normal of the waveguide and the gravity direction is $\sim5$~mrad.

\begin{figure*}
\includegraphics[width=1\textwidth]{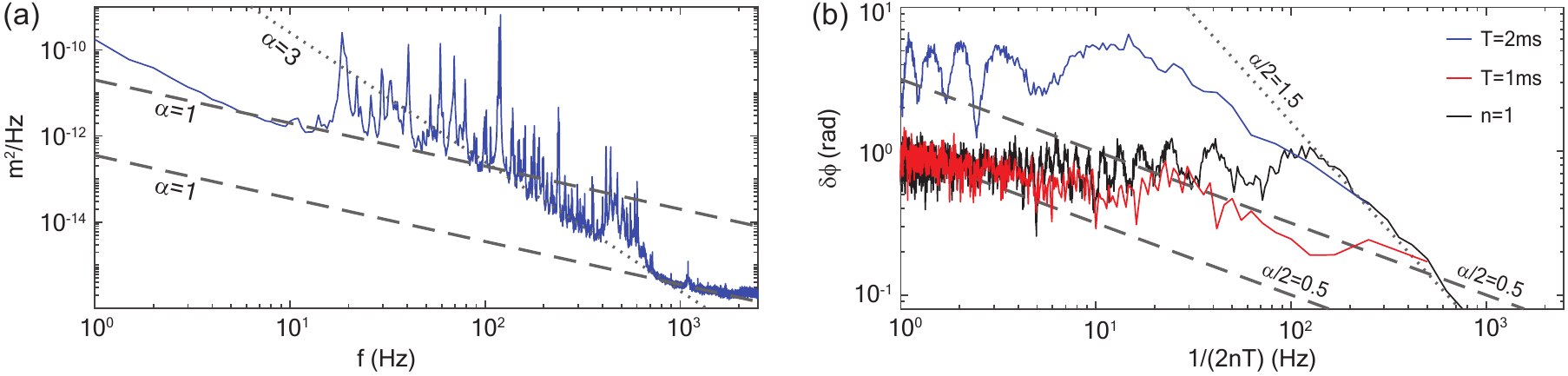} 
\caption{\label{fig:MirrorNoiseSpectrum} The mirror position noise spectrum and the corresponding phase noise.  (a) The blue line is the reference mirror position noise spectrum density, measured by the independent optical Michelson interferometer. (b) Color lines are the phase fluctuation $\delta\phi_{laser}=\sqrt{\sum_i\delta\phi_{laser}^{i,2}}$ in Eq.~(\ref{eq:PhaseFluctuation}). Gray dashed and dotted lines are the noise slope guide.}
\end{figure*}

The implications for the interferometer sequence are as follows. The estimated tilt, which is ($\theta_{tilt}\ll 1$), does not affect the phase of the atom interferometer for $n=even$ as long as the magnetic field is static. The excursion of each wavepacket ($\pm30~\mu$m at $T=1.125~$ms) is longer than the wavepacket size $\sigma_z=6\sim20~\mu$m determined by a Gaussian density fit $|\psi|^2\propto\exp(-z^2/2\sigma_z^2)$ (The wavepacket size is time-dependent due to the compressional mode). Therefore the wave packets are well-separated from each other during the interferometer cycle. Finally, the ratio of compressional mode frequency and the axial trap frequency $\omega_c/\omega_z=[2.1(2),1.8(1),1.7(2)]$ for the three scattering lengths shows that our system lies between the non-interacting and 1D Thomas-Fermi regimes~\cite{dicarli2019excitation, haller2009realization}. 

\section{Phase noise due to retro-reflecting mirror vibration}
The interferometer phase ($\phi_{laser}$) and the phase fluctuation due to technical noise ($\delta\phi_{laser}^i$) in shot $i$ are given by
\begin{widetext} 
\begin{equation}
\label{eq:LaserPhase}
\begin{split}
 \phi_{laser}=  &4k_0z(0)+8k_0\sum_{j=1}^{j=n}(-1)^jz(t_j)+4k_0(-1)^{n+1}[z(t_{n+1})+\frac{\theta}{4k_0}], \\
 \delta \phi_{laser}^i=&4k_0\delta z^i(0)+8k_0\sum_{j=1}^{j=n}(-1)^j\delta z^i(t_j)+4k_0(-1)^{n+1}\delta z^i(t_{n+1}),
\end{split}
\end{equation}
\end{widetext}
where $z(t_j)$ and $\delta z^i(t_j)$ are respectively the center-of-mass position of the atoms and the displacement of the reference mirror when the $j$-th pulse hits. Here, $k_0$ is the wavevector of the Bragg light. Note that the $0$-th and $n+1$-th pulses are splitting pulses.  To find $z(t_j)$, we used a 4-th order Runge-Kutta method to numerically calculate the trajectory of the atoms using $F=ma$ including the known guide potential and assuming that the photon momentum transfer is instantaneous. To obtain $\delta z^i(t_j)$, we independently measured the reference mirror vibration with an optical Michelson interferometer. The resulting noise amplitude spectrum is shown in the inset of Fig.~\figFringes(b) with a fit to the function $1/f^{\alpha/2}$. For the numerical estimation of the shot-to-shot phase noise $\phi_{laser}^i$, $\delta z^i(0)$ is randomly sampled from the time series of the Michelson interferometer data.

\begin{figure}
\includegraphics[width=1\columnwidth]{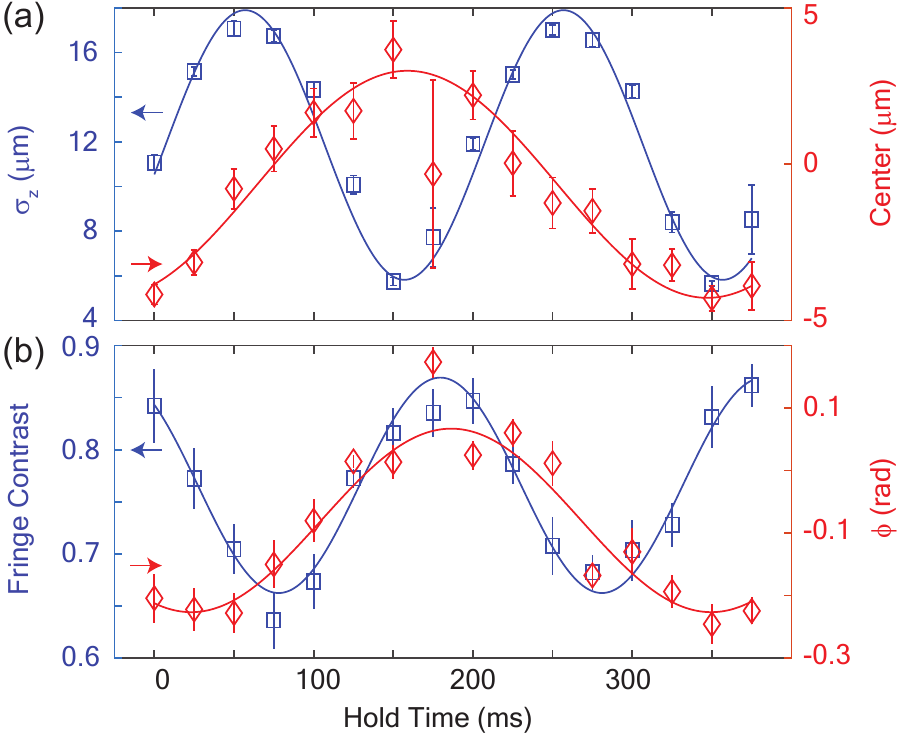} 
\caption{\label{fig:nEqualsOne} $n=1$ interferometer. (a) $\sigma_z$ (left) and center-of-mass position (right), and (b) fringe contrast (left) and phase (right) are presented as a function of hold time for $T=2.125~$ms, $a_s=1.7a_0$. Solid lines are sinusoidal fits. The oscillations in $\sigma_z$ and fringe contrast (center and $\phi$) are out-of (in)-phase, with a small shift due to the time-of-flight. Error bars are $1\sigma$.}
\end{figure}
 
\section{Noise spectrum and phase noise}
The main sources of reference mirror vibration are the flow of cooling water inside the magnetic field coils and movement of the floor supporting the optical table. Note that the mirror vibration measured by the optical Michelson interferometer is re-scaled by $0.38$ to fit the experimental data in Fig.~\figFringes(b). In this section, we focus on the reference mirror vibration-induced phase noise and its relation to the noise spectrum. From Eq.~(\ref{eq:LaserPhase}), the average laser phase fluctuation becomes

\begin{equation}
\label{eq:PhaseFluctuation}
\delta\phi_{laser}^2\propto\overline{\delta z^i(t_{n+1})^2}+2\sum_{j=1}^{j=n}\overline{\delta z^i(t_j)^2},
\end{equation}
when the displacement of the reference mirror $\delta z^i(t)$ is assumed to be a random walk-type fluctuation. Note that $\delta z^i(0)\equiv 0$ because the displacement at the start of the cycle is taken to be zero. If $\delta z(t)^2$ is Markovian noise with spectrum $1/f^\alpha$, the average value $\overline{\delta z^i(t_j)^2}\propto(t_j-t_{j-1})^\alpha$ can be deduced. In the case of a multiple-loop interferometer with parameters $n$ and $T$, Eq.~(\ref{eq:PhaseFluctuation}) becomes $\delta\phi^2\propto3(T)^\alpha+2(n-1)(2T)^\alpha\approx2n(2T)^\alpha$ for $n\gg1$. The corresponding result for a single-loop interferometer with the same interrogation time is $\delta\phi^2\propto3(nT)^\alpha$. Consequently in the case $n\gg1$ and $\alpha>1$ the phase noise of the multiple-loop scheme can be smaller than that of the single-loop scheme, as seen in Fig.~\ref{fig:MirrorNoiseSpectrum}(b) by comparing the red and black curves.

The phase noise calculated using  Eq.~(\ref{eq:LaserPhase}) applied to the optical interferometer data for $T=1~$ms (red curve in Fig.~\ref{fig:MirrorNoiseSpectrum}(b)) has scaling exponent close to $0.5$ around $100~$Hz, and it shows good agreement with the scaling of $\sqrt{n}$ from Eq.~(\ref{eq:PhaseFluctuation}). The phase noise for $n=1$ (black line) has scaling exponent close to $1.5$ around $100~$Hz, which resembles the slope of the noise density in Fig.~\ref{fig:MirrorNoiseSpectrum}(a) and agrees with Eq.~(\ref{eq:PhaseFluctuation}). The phase noise for $T=2~$ms (blue line) also has a similar slope, which explains the rapid loss of the contrast in Fig.~\figFringes(b).

The signal-to-noise ratio (SNR) scales as $4k_0a_{max}(nT)^{2-\alpha/2}/\sqrt{3}$ for a single-loop scheme measuring a constant uniform acceleration $a_{max}$, when $\delta\phi_{laser}^2$ is the dominant noise. The SNR for a multiple-loop scheme measuring a square-modulated acceleration scales as $2^{(1-\alpha)/2}k_0a_{max}n^{1/2}T^{2-\alpha/2}$. Therefore the multiple-loop scheme has better SNR when the slope $\alpha$ of the phase noise satisfies $\alpha>3$.
 
\section{Numerical simulation of fringe contrast}
Equation~(\ref{eq:LaserPhase}) treats the atoms as classical point-like particles. For a more complete simulation using finite size wave packet, the maximum fringe contrast is $C_{max}=\int |\psi_+\psi_-^*|\cos\phi_{laser}(z)dz$, where $|\psi_+\psi_-^*|$ is the probability distribution of combined wavepacket and $\phi_{laser}(z)$ is the axial position dependent phase from Eq.~(\ref{eq:LaserPhase}). The averaged fringe contrast is $C=C_{max}\times\overline{\cos}\delta\phi_{laser}^i$. Note that $\phi_{laser}(z)=4k_0z(\omega_zT)^2$ for $n=odd$ and $\omega_zT\ll1$, which is an inhomogeneous dephasing imprinted along the wave packet due to the axial curvature of the potential. Note that if $n$ is even then the inhomogeneous dephasing terms cancel to zero, leaving the next highest order contribution as the leading term. Eq. 26 in Reference~\cite{burke2016confinement} has shown that the next order is $(\omega_zT)^4$, which agrees with our numerical estimation of $\phi(z)=\phi_{laser}+\phi_{path}\sim 47k_0z(\omega_zT)^4$, where $\phi_{path}=(S_+-S_-)/\hbar$ is the phase from the classical action~\cite{haslinger2018attractive}; the second order term comes from the energy loss when the atoms climb up the potential while the momentum kicks are always constant.
 
The blue dashed line in Fig.~\figInterrogationTimeLimits(a) is calculated for $C_{max}=\int |\psi_+\psi_-^*|\cos(\phi(z))dz$, where $\psi_{\pm}=\exp[-(z\pm\Delta z)^2/4\sigma_z^2]$, $\phi(z)=47k_0z(\omega_zT)^4$, and $\sigma_z=14~\mu$m is the average size over the hold time. The center-of-mass shift due to the potential curvature is numerically found to be $\Delta z\propto T^2\omega_zk_0\sin{(2n\omega_zT)}$, which is negligible compared to $\sigma_z(t)$. When the time-varying $\sigma_z$ due to the compressional mode of the BEC is considered, $\phi(z)=4k_0(\omega_zT)^2z\times[\sigma_z(4T)/\sigma_z(0)-1]$, becoming effectively proportional to $k_0z(\omega_zT)^3$ as the inhomogeneous dephasing for $n=1$ is only partially canceled out. The blue solid line in Fig.~\figFringes(a) takes this into account and shows good agreement with our experiment.

\section{\textit{n}=1 and \textit{n}=2 interferometer}
Fig.~\ref{fig:nEqualsOne} shows for $n=1$ the axial width $\sigma_z$ and center-of-mass position (a), and the fringe contrast $C$ and phase $\phi$ (b). The amplitude of the phase shift calculated from the center-of-mass oscillation is $\phi_{max}=4k_0a_{max}T^2=0.17(1)$~rad, where $a_{max}=\omega_z^2z_{initial}$ and $z_{initial}=3.6(3)~\mu$m. The result shows good agreement with the actual interferometer phase measurement $\phi_{max}=0.15(2)~$rad (red points in Fig.~\ref{fig:nEqualsOne}(b)). The interferometer phase offset from $0$ might be attributable to the asymmetry of the combined waveguide and magnetic potential. 

\begin{figure*}
\includegraphics[width=1\textwidth]{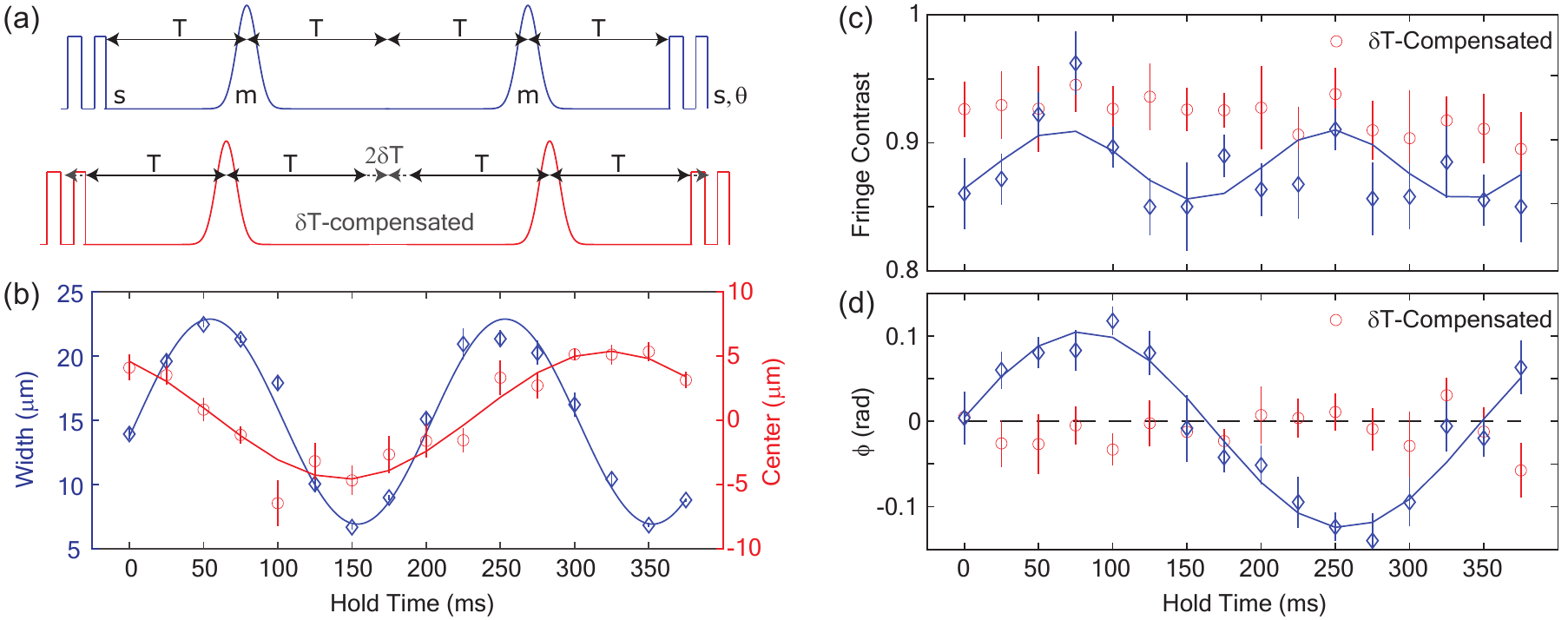} 
\caption{\label{fig:nEqualsTwo} $n=2$ interferometer. (a) The pulse timing that compensates the finite splitting pulse length (bottom), compared to a careless timing (top). (b) The axial width $\sigma_z$ and center-of-mass position as a function of hold time. The fringe contrast (c) and the phase (d) for $n=2$, $T=1.125~$ms, $a_s=1.7a_0$. The splitting pulse compensation (red circles) and the other (blue diamonds) are compared. Error bars are $1\sigma$. The fringe contrast is modulated in-phase to the axial width, and the interferometer phase is modulated $90^\circ$ in advance to the center-of-mass.}
\end{figure*}

The measured fringe contrast (blue points in Fig.~\ref{fig:nEqualsOne}(b)) is out-of-phase with the oscillation of the compressional mode (blue points in Fig.~\ref{fig:nEqualsOne}(a)) as the inhomogeneous dephasing $\phi(z)=4k_0z(\omega_zT)^2$ exacerbates the fringe contrast degradation for larger wavepacket size. The wavepacket does not perfectly overlap after the interrogation due to the potential curvature. However, the non-overlapped portion is less than a few percent compared to the wavepacket size for $T \leq 2~ms$.

Figure~\ref{fig:nEqualsTwo} presents the corresponding set of data for $n=2$, $T=1.125~$ms, $a_s=1.7a_0$. In the case of $n=2$, the inhomogeneous dephasing and DC acceleration sensitivity for $n=1$ are canceled out by the conjugate path. We observed the modulation of fringe contrast and momentum dependent phase shift (blue diamonds in Fig.~\ref{fig:nEqualsTwo}(c) and (d), respectively) when the pulse time interval did not compensate for the finite length of the splitting pulses (the first scheme in Fig.~\ref{fig:nEqualsTwo}(a)). This effect can be clarified by solving the Schr\"{o}dinger equation,
\begin{equation}
\label{eq:Schrodinger}
i\frac{d\psi}{dt}=\left[\frac{\hbar}{2m}\frac{d^2}{dz^2}+\Omega(t)\cos(2k_0z+\theta/2)\right] \psi,
\end{equation}
where $\Omega(t)$. Expanding the wave function in the Bloch basis and substituting it in Eq.~(\ref{eq:Schrodinger}) gives
\begin{widetext}
\begin{equation}
\label{eq:PlaneWaveExpansion}
i\frac{dC_{2j}(k,t)}{dt}=\frac{\hbar}{2m}(2jk_0+k)^2C_{2j}(k,t)+\frac{\Omega(t)}{2}\left[C_{2j-2}(k,t)e^{i\theta/2}+C_{2j+2}(k,t)e^{-i\theta/2}\right].
\end{equation}
\end{widetext}

After taking the initial condition to be atoms at rest and assuming a perturbative regime in which $C_{2j}(k,0)=\delta_{j,0}f(k)$ and $\Omega(t)\ll32\omega_r$, where $|f(k)|^2$ is a narrow ($\Delta k\ll k_0$) momentum distribution centered at $0$ and $\omega_r=\hbar k_0^2/2m$, Eq.~(\ref{eq:PlaneWaveExpansion}) is truncated to the lowest 3 states,

\begin{widetext}
\begin{equation}
\begin{split}
\label{eq:ThreeStateTruncation}
i\dot{C_0}&=\frac{\Omega(t)}{2}(C_{-2}+C_{2})\cos\frac{\theta}{2}+i\frac{\Omega(t)}{2}(C_{-2}+C_2)\sin\frac{\theta}{2}, \\
i\dot{C_2}&=4\omega_r\left(1+\frac{k}{k_0}\right)C_2+\frac{\Omega(t)}{2}e^{i\theta/2}C_0, \\
i\dot{C_{-2}}&=4\omega_r\left(1-\frac{k}{k_0}\right)C_{-2}+\frac{\Omega(t)}{2}e^{-i\theta/2}C_0.
\end{split}
\end{equation}
\end{widetext}

Note that the kinetic energy associated with $k$ has been subtracted by a unitary transformation. Eq.~(\ref{eq:ThreeStateTruncation}) is numerically solved with an appropriate potential shape $\Omega(t)$ as in Fig.~\ref{fig:nEqualsTwo}(a). In Eq.~(\ref{eq:ThreeStateTruncation}) $\theta$ is the relative phase between the recombination and the splitting pulses, so $\theta$ is set to $0$ for all the pulses except the recombination pulse. The initial center-of-mass momentum $k$ in Eq.~(\ref{eq:ThreeStateTruncation}) can cause a phase shift between the left and right atoms at the time of recombination. This happens when the amount of time the atoms spent in the bases $C_{-2}$ and $C_{2}$ are not equal. The timing compensation $\delta T=22~\mu$s determined numerically was found experimentally to optimize the fringe contrast and to flatten the momentum-dependent phase shift as in Fig~\ref{fig:nEqualsTwo}(c) and (d). Note that the compensation is valid until $\hbar k <0.05\hbar k_0$, which is larger than the initial momentum of our BEC ($<0.01\hbar k_0$) induced by the loading process.

\section{Scaling behavior of cold collisions}

\begin{figure}
\includegraphics[width=\columnwidth]{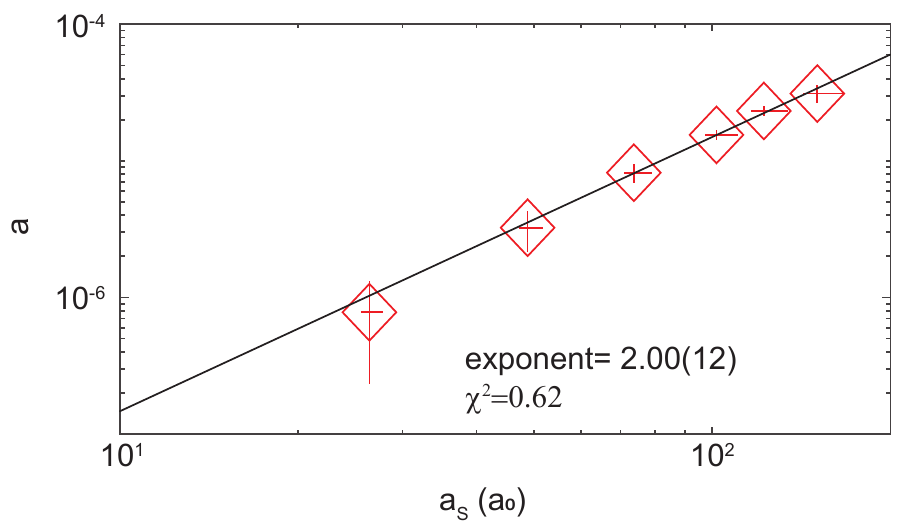} 
\caption{\label{fig:LogLogPlot} The two-body loss coefficient $a$ as a function of s-wave scattering length $a_s$ in log-log plot. The black solid line is the least-square fit, extracting the scaling exponent. Error bars are $1\sigma$. The s-wave scattering length is $a_s=-29a_0\times[1+56~\text{G}/(B-562.2~\text{G})]$, and the $B$ field is independently measured by RF spectroscopy of the atoms.}
\end{figure}

The multiple-loop atom interferometer suffers from cold collision-induced atom loss~\cite{band2000elastic} that can usually be neglected for a small number of loops. As shown in Fig.~\figExperimentalScheme(a), the atomic wave packets pass through each other once for every mirror pulse, and so the chance of a collision scattering an atom out of the coherent wave packet is $n$ times larger than the $n=1$ single reflection scheme. It results in a collisional loss that scales as $\text{d}N_{tot}/\text{d}n\propto -a_s^2N_{tot}^2$, where $N_{tot}, n$, and $a_s$ are the atom number, the number of collisions, and the s-wave scattering length, respectively. As shown in Fig.~\figInterrogationTimeLimits(b), we measured the atom number as a function of collision number for various s-wave scattering lengths. The fitted curves are the analytic solution of the  differential equation $\text{d}N_{tot}/\text{d}n=-aN_{tot}^2-bN_{tot}$, which is $N_{tot}(n)=-b\exp(bc)/[a\exp(bc)-\exp(bn)]$, where $c$ is a constant related to the initial atom number and $n$ is treated as a continuous variable for simplicity. The results of the least square fits are listed in Table~\ref{tab:Fit}. From the results, the two-body loss coefficient $a$ vs. s-wave scattering length $a_s$ is plotted in log-log scale in Fig.~\ref{fig:LogLogPlot} and the scaling exponent is extracted as $2.00(12)$. Note that the first two data points are not shown on the plot. We found that over the wide range of s-wave scattering length $a_s=-2\sim150a_0$ fringe contrast remained flat with $C>0.7$ regardless of $n$, as long as atom numbers were well above the detection noise. However, we emphasize that tuning the s-wave scattering length to near zero is of essential to obtain atomic signal for larger number of loop $n>5$ (c.f. Fig.~2(b)). That result suggests that dephasing from the interatomic interaction is negligible compared to the collisional loss for our interrogation time-scales and atomic density once the s-wave scattering length is set to near zero. Note that the initial mean atomic density at $a_s=1.7a_0$ is estimated as $2.9\rm E13-4.5E13~/cm^3$, depending on the initial loading condition.

The mirror pulse efficiency $1-b$ has the highest value at $a_s<6a_0$, and gradually reduces at larger $a_s$ because of the enlarged interaction-induced momentum width~\cite{schubert2021multi,fattori2008atom,roati200739K}. Note that the initial center-of-mass momentum also degrades the mirror efficiency but the theoretical upper bound for our case is $99.98\%$, which is negligible degradation. However, the momentum width from the compressional mode currently limits the mirror efficiency to $99.4\%$. An atom number $N_{tot}\sim800$ remains after $n=400$ reflections with the current mirror efficiency, which is comparable to the atom and photon shot noise of $150$ atoms.

\begin{table}
\caption{The results of the least square fit.}
\begin{ruledtabular}
\begin{tabular}{ccccc}                              
  \multicolumn{1}{c}{$a_s(a_0)$} 
& \multicolumn{1}{c}{$a$}  
& \multicolumn{1}{c}{$b$} 
& \multicolumn{1}{c}{$c$} 
& \multicolumn{1}{c}{$\xi^2$} \\
1.7        &  \multicolumn{1}{c}{$2.42(0)\times 10^{-9}$}   & 0.0065(5)   & 2160(163)   & 8.36  \\
14.0       &  \multicolumn{1}{c}{$7.96(0)\times 10^{-9}$}   & 0.0106(3)   & 1290(42)    & 4.97  \\
26.4       &  \multicolumn{1}{c}{$7.8(5.5)\times 10^{-7}$}  & 0.0127(24)  & 1020(186)   & 9.11  \\
48.8       &  \multicolumn{1}{c}{$3.24(55)\times 10^{-6}$}  & 0.0151(15)  & 811(77)     & 7.57  \\
73.7       &  \multicolumn{1}{c}{$8.19(64)\times 10^{-6}$}  & 0.0154(13)  & 751(58)     & 5.34  \\
102.0      &  \multicolumn{1}{c}{$2.32(10)\times 10^{-5}$}  & 0.0143(13)  & 741(62)     & 6.10  \\
150.4      &  \multicolumn{1}{c}{$3.11(24)\times 10^{-5}$}  & 0.0161(23)  & 640(88)     & 14.6 
\end{tabular}
\end{ruledtabular}
\label{tab:Fit}
\end{table}

\begin{figure*}
\includegraphics[width=1\textwidth]{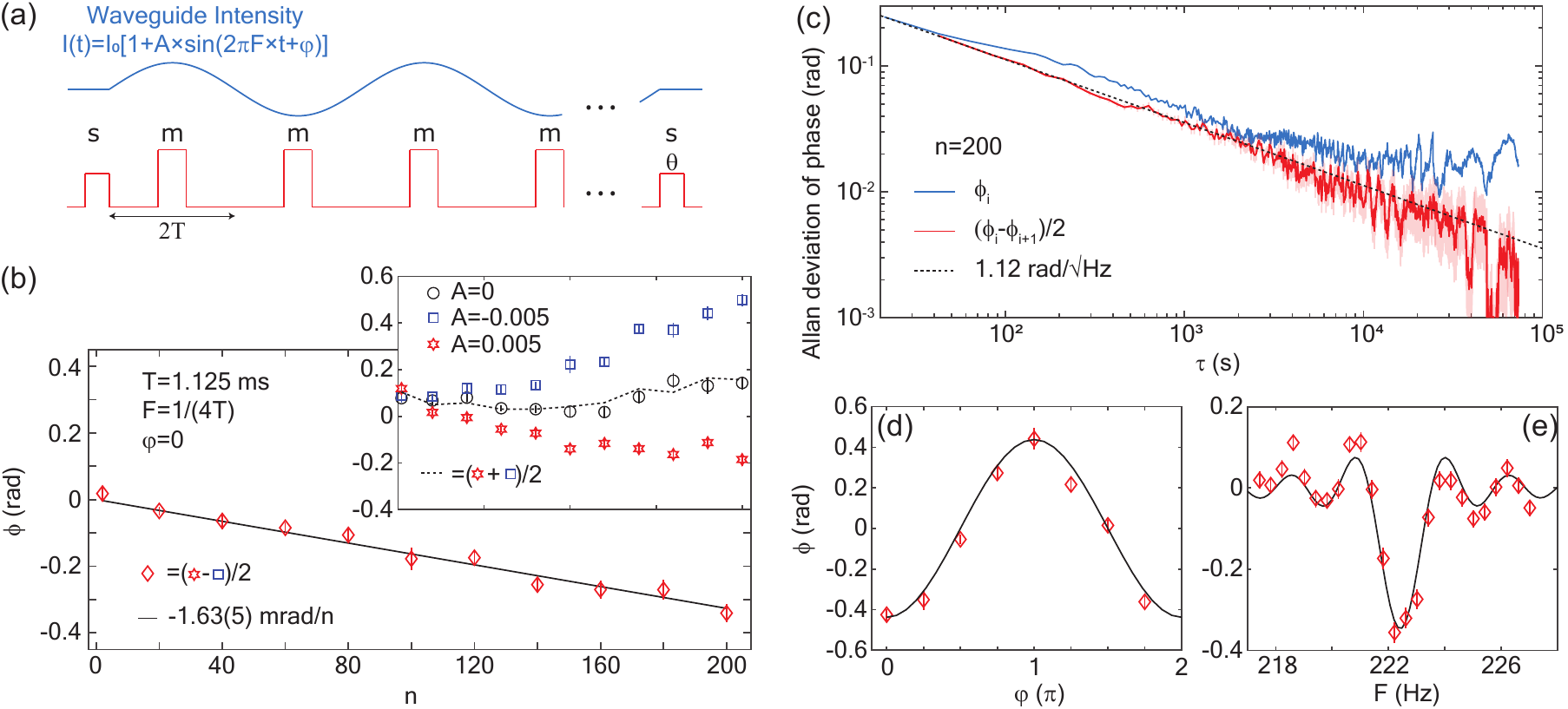} 
\caption{\label{fig:QuantumLockin} Quantum lock-in amplification of the atom interferometer. (a) The scheme for phase-synchronous signal measurement. The intensity of the waveguide is modulated. (b) Linear amplification of the interferometer phase $\phi$ as a function of $n$ for a modulated waveguide intensity. Symbols are odd-parity phases. The line is a linear fit. Up to 20 dB signal amplification is demonstrated. (inset) Symbols are $\phi$ vs. $n$ for different modulation amplitudes. The black dotted line is even-parity phase that corresponds to systematic phase shift at $A=0$. (c) Allan deviation of $\phi$ for raw data (blue) and odd-parity of successive phases (red) for $n=200$. The odd-parity operation reduces systematic shift and long-term drift. (d) $\phi$ (symbols) as a function of $\varphi$ for fixed $A$ and $F=1/(4T)$. The black line is a sine curve fit. The presented $\phi$ is offset by the systematic phase. (e) Odd-parity phases (symbols) vs. $F$ and $A=0.005$. The black line is from lock-in amplifier theory. $\phi$ is measured by either scanning $\theta$ [(b),(d)] or fixing $\theta$ at the steepest slope [(c),(e)]. Error bars are the standard error in the mean.}
\end{figure*}

\section{Quantum lock-in amplification of atom interferometer}
The multiple-round trip scheme allows for phase-synchronous measurement of a suitably-modulated perturbation, a realization of quantum lock-in amplification \cite{kotler2011single}. To make a proof-of-principle demonstration of this capability, the waveguide intensity was modulated as  $I(t)=I_0[1+A\sin(2\pi F t+\varphi)]$. Note that since the initial atom position is displaced from the waveguide waist position, the waveguide beam intensity modulation induces a modulated acceleration $a(t)=a_{max} A\sin(2\pi F t+\varphi)$. The laser phase is $\phi_{laser}=8k_0\sum_{j=1}^{j=n}(-1)^ja_{max}A/(2\pi F)^2=32k_0a_{max}T^2n/\pi^2$ for $F=1/(4T)$, and $\varphi=0$. From the data of Fig.~\ref{fig:QuantumLockin}(b), we estimated $a_{max}=4.85(15)\times10^{-5}~\text{m/s}^2$. Note that the odd-parity operation of successive shots of alternating $\varphi=\{0,\pi\}$ (i.e. differential average) mitigates the systematic phase shift. We estimate the beam waist displacement $z$ that satisfies $a_{max}=-\mathrm{d}U(z)/(m \mathrm{d}z)$, where $U(z)=U_0w_0/[1+(z/z_R)^2]$, $U_0=4.6~\mu$K, $w_0=100~\mu$m, and $z_R=\pi w_0^2/1064$~nm. The beam waist location is estimated as $4.5~$mm away from the atom position, which is smaller than the Rayleigh length $30~$mm.

Figure~\ref{fig:QuantumLockin}(c) shows the Allan deviations of the raw phase and the odd-parity phase, respectively. The odd-parity operation improves long-term stability as the systematic phases were canceled out. The black dotted line consists of $30\%$ atom and photon shot noise, and $70\%$ mirror vibration and laser phase noise.

Figure~\ref{fig:QuantumLockin}(d) and (e) present the response of lock-in amplification as a function of $\varphi$ and $F$, respectively. The theory curves are from Eq.~(\ref{eq:LaserPhase}). The theoretical gain of the lock-in amplifier is $\log_{10}(n/2)~$dB. When the system is atom shot-noise limited, it has maximum SNR at $n=\sqrt{1/b}$.



\end{document}